\documentclass[structabstract]{aa}  

\usepackage{graphicx}
\usepackage{graphicx,color,layout}
\usepackage{epstopdf}
\usepackage{txfonts}
\usepackage{natbib}
\usepackage{lscape}
\usepackage{longtable}
\usepackage{threeparttable}
\usepackage[dvipsnames]{xcolor}

\newcommand\parametersstep{140}

\newcommand\allnpf{192}

\newcommand\binarynpf{12}
\newcommand\rotatenpf{71}
\newcommand\parametersnpf{109}

\newcommand\allfull{405}

\newcommand\parametersfull{249}
\newcommand\numthin{223}
\newcommand\numthick{26}
\newcommand\numrvincommon{153}

\defcitealias{Mikolaitis2018}{Paper~I}

\begin{document}
   \title{High-resolution spectroscopic study of dwarf stars \\ in the northern sky 
}
   \subtitle{Na to Zn abundances in two fields with radii of 20 degrees
   \thanks{Based on observations collected with the 1.65~m telescope and VUES spectrograph at the Mol\.{e}tai Astronomical Observatory of Institute of Theoretical Physics and Astronomy, Vilnius University, for the SPFOT survey.}
\thanks{
Full Tables~\ref{tab:TABLE_LINES}~and~\ref{tab:CDS} are only available in electronic form at the CDS via anonymous ftp to cdsarc.u-strasbg.fr (130.79.128.5) or via http://cdsweb.u-strasbg.fr/cgi-bin/qcat?J/A+A/}
   }

   \author{
            \v{S}. Mikolaitis
          \inst{1},
            A. Drazdauskas
          \inst{1},
          R. Minkevi\v{c}i\={u}t\.{e}
          \inst{1},
          E. Stonkut\.{e}
          \inst{1},
            G. Tautvai\v{s}ien\.{e}
          \inst{1},
            L. Klebonas
          \inst{1,2}, 
          V. Bagdonas
          \inst{1} \\
	  E. Pak\v{s}tien\.{e}
          \inst{1},
\and
          R. Janulis
          \inst{1}          
                 }

   \institute{Astronomical Observatory, Institute of Theoretical Physics and Astronomy, Vilnius University, Saul\.{e}tekio av. 3, 10257 Vilnius, Lithuania\\
              \email{Sarunas.Mikolaitis@tfai.vu.lt} \\
              \and
              Mathematisch-Naturwissenschaftliche Fakult\"at, Universit\"at Bonn, Wegelerstra{\ss}e 10, 53115 Bonn, Germany
             }
   \date{Received 31 December 2018/ Accepted 10 May 2019}

\titlerunning{High-resolution spectroscopic study of dwarf stars in the northern sky } 
\authorrunning{\v{S}. Mikolaitis et al.}

  \abstract
  % context heading (optional)
{New space missions, such as NASA TESS or ESA PLATO, will focus on bright stars, which have been largely ignored by modern large surveys, especially in the northern sky. Spectroscopic information is of paramount importance in characterising the stars and analysing planets possibly orbiting them, and in studying the Galactic disc evolution. 
   }
  % aims heading (mandatory)
{The aim of this work was to analyse all bright ($V < 8$~mag)  F, G, and K dwarf stars using high-resolution spectra in the selected sky fields near the northern celestial pole.
}
  % methods heading (mandatory)
{%We have observed high-resolution spectra of \parametersfull ~F, G, and K dwarf stars in the northern sky up to the V = 8 mag. 
The observations were carried out with the 1.65~m diameter telescope at the Molėtai Astronomical Observatory and a fibre-fed high-resolution spectrograph covering a full visible wavelength range (4000--8500~$\AA$). The atmospheric parameters were derived using the classical equivalent width approach while the individual chemical element abundances were determined from spectral synthesis. For both tasks the one-dimensional plane-parallel LTE MARCS stellar model atmospheres were applied.
The NLTE effects for the majority of elemental abundances in our sample were negligible; however, we did calculate the NLTE corrections for the potassium abundances, as they were determined from the large 7698.9~$\AA$ line. For manganese and copper we have accounted for a hyperfine splitting. 
}
  % results heading (mandatory)
{We determined the main atmospheric parameters, kinematic properties, orbital parameters, and stellar ages for \parametersnpf~newly observed stars and chemical abundances of \ion{Na}{I}, \ion{Mg}{I}, \ion{Al}{I}, \ion{Si}{I}, \ion{Si}{II}, \ion{S}{I}, \ion{K}{I}, \ion{Ca}{I}, \ion{Ca}{II}, \ion{Sc}{I}, \ion{Sc}{II}, \ion{Ti}{I}, \ion{Ti}{II}, \ion{V}{I}, \ion{Cr}{I}, \ion{Cr}{II}, \ion{Mn}{I}, \ion{Fe}{I}, \ion{Fe}{II}, \ion{Co}{I}, \ion{Ni}{I}, \ion{Cu}{I}, and  \ion{Zn}{I} for \parametersfull~ F, G, and K dwarf stars observed in the
present study and in our previous study.  The [\ion{Mg}{I}/\ion{Fe}{I}] ratio was adopted to define the thin-disc ($\alpha$-poor) and thick-disc ($\alpha$-rich) stars in our sample. We explored the behaviour of 21 chemical species in the [El/\ion{Fe}{I}] versus [\ion{Fe}{I}/H] and [El/\ion{Fe}{I}] versus age planes, and compared the results with the latest Galactic chemical evolution models. We also explored [El/\ion{Fe}{I}] gradients according to the mean Galactocentric distances and maximum height above the Galactic plane.}
{
We found that in the Galactic thin-disc [El/\ion{Fe}{I}] ratios of $\alpha$-elements and aluminium have a positive trend with respect to age while the trend of Mn is clearly negative. Abundances of other species do not display significant trends. While the current theoretical models are able to reproduce the generic trends of the elements, they often seem to overestimate or underestimate the observational abundances. We found that the $\alpha$-element and zinc abundances have slightly positive or flat radial and vertical gradients, while gradients for the odd-{\it Z} element Na, K, V, and Mn abundances are negative. }

   \maketitle

\section{Introduction}
\label{sec:intro}

In the era of the new extra-solar planet hunting telescopes such as   ESA's PLAnetary Transits and Oscillations of stars (PLATO) or NASA's  Transiting Exoplanet Survey Satellite (TESS), it is very important to obtain as much information as possible on the stars they will be observing. These missions in particular will focus on the bright targets (\textit{V} < 12~mag), that are perfectly suitable for observations with smaller ground-based telescopes. Only less than 30\% of bright dwarf stars in the solar neighbourhood have been studied spectroscopically. 
There are more than a few spectroscopic surveys currently in progress, $Gaia$-ESO \citep{2012Msngr.147...25G}, GALAH \citep{Zucker12}, APOGEE \citep{2017AJ....154...94M}, among others; however, these surveys are either based in the southern hemisphere or have other limitations such as limited wavelength coverage or a focus on fainter stars.

Coupled with the fact that the brightest stars tend to be ignored by recent surveys, we wanted to fill the existing data gap, and so have started a project to observe the bright dwarf stars in the northern sky (see \citealt{Mikolaitis2018} for the overview of the project, hereafter \citetalias{Mikolaitis2018}).

The position of our telescope (55$^\circ$18$^\prime$ northern latitude) gives us a unique opportunity to comfortably observe stars around the northern celestial pole. Thus, for our project we employed the 1.65~m telescope at the Molėtai Astronomical Observatory of Vilnius University together with the high-resolution Vilnius University Echelle Spectrograph (VUES) \citep{Jurgenson2016} covering the whole visible wavelength range (4000 -- 8500~\AA). With the data from VUES we were able to determine the main atmospheric parameters of stars ($T_{\rm eff}$, log\,\textit{g}, [Fe/H]) and a detailed chemical composition. 

We started observations in two sky-fields with radii of 20 degrees located close to the northern celestial pole. 
In \citetalias{Mikolaitis2018} we presented the atmospheric parameters for \parametersstep~ stars in one of the fields. In this paper we add results that include the main atmospheric parameters for stars in the second field, as well as detailed kinematic parameters and elemental abundances for stars in both fields.

It is already known that our Galaxy contains distinct disc components, notably the thin- and thick-discs. However, the exact mechanism through which these two populations formed is still debatable (e.g. \citealt{Grand2018} and references therein). Studying the detailed chemical composition and kinematic properties for new samples of stars
can yield deeper
insights into the Galaxy formation. 

The solar neighbourhood contains stars from all local Galactic components (thin- and thick-discs, and the halo); however, the majority of stars belong to the thin-disc. It is known that the thick-disc  generally contains older and more metal-poor stars compared to the thin-disc, with different orbital parameters,  longer scale length and larger scale height (\citealt{Kordopatis2011}). The most prominent feature that separates the Galactic components is  the different chemical composition,  the ratio of  $\alpha$-elements to iron  in particular (see \citealt{Grisoni2017} and references therein).
The solar neighbourhood has been the focus of a number of recent studies (\citealt{Bensby2014,Duong2018,Frasca2018,Mishenina2017,Feuillet2018} among others). These studies differ in  the  stars  selected,  how they were selected, which elements were analysed, and how detailed  the analysis was. 

In this study we explore the radial and vertical abundance gradients for 21 chemical species and 
compare our results with the latest Galactic chemical evolution models (\citealt{Romano2010, Kobayashi2011, Prantzos2018}). Even though the investigated stars are not spread over large distances, as this is a study of the solar neighbourhood, if we involve the orbital parameters ($z_{max}$ and $R_{\mathrm{mean}}$), metallicity, and age, we can investigate the original spatial distribution of the stars and study the Galactic elemental abundance gradients successfully. Our study also uses the data from the European Space Agency mission Gaia \citep{Gaia2018}.

This paper is organised as follows. In Sect.~\ref{sec:targets} we describe the observational data. In Sect.~\ref{sec:methods} we present the methods of analysis used to determine stellar radial velocities, kinematic properties, ages, and chemical abundances, as well as possible uncertainties. In Sect.~\ref{sec:atmospheres} we describe the derived stellar atmospheric parameters. In Sect.~\ref{sec:kinematicparameters} we discuss kinematic parameters and ages of the investigated stars. In Sect.~\ref{sec:ratios} we address the determined element abundance ratios and their comparison with the Galactic evolution models. In Sect.~\ref{sec:gradients} we study the radial and vertical abundance gradients in the Galactic thin-disc. In Sect.~\ref{sec:summary} we summarise the work and the results.
 
\section{Target selection, observations, and data processing}
\label{sec:targets}

\begin{figure}[]
   \advance\leftskip 0cm
   \centering
   \includegraphics[width=\columnwidth]{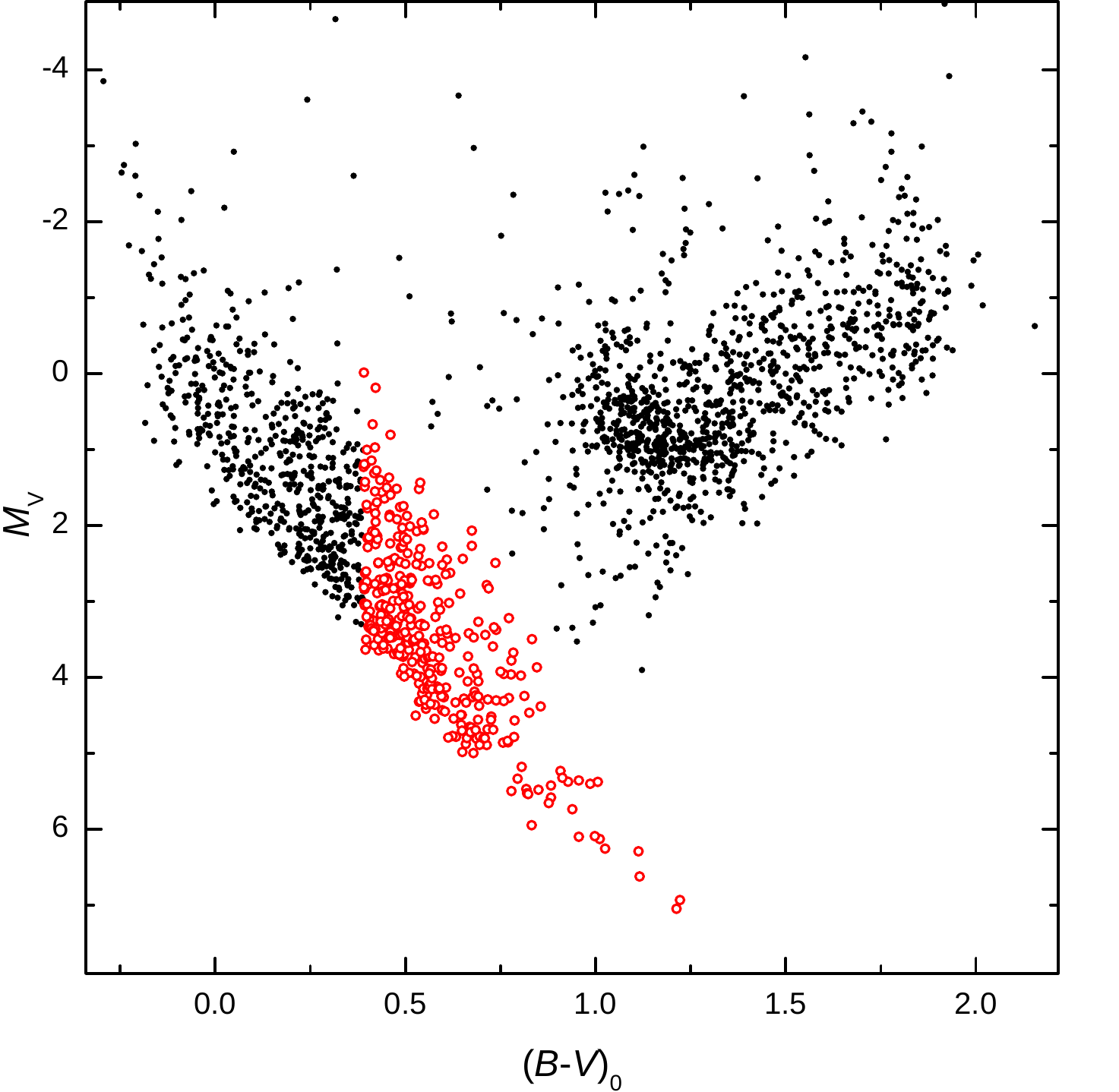}
  \caption{Colour-magnitude diagram of stars $V < 8$~mag in two fields. The FGK dwarfs (\allfull~stars) observed in this programme are presented as red open circles.}
  \label{fig:cmd_targets}
  \end{figure}

The methods we used for the target selection, observations, and data processing are described in \citetalias{Mikolaitis2018}. Here we only briefly summarise the most important aspects of this work.
In \citetalias{Mikolaitis2018} we describe our study of the dwarf stars situated around the centre of the preliminary PLATO STEP02 field (20~degrees around $\alpha$(2000)~=~161.03552$^\circ$ and $\delta$(2000)~=~86.60225$^\circ$, hereafter the first field; see \citealt{Rauer2014,Rauer2016}). For the second step of our project we observed another field of the northern sky of the same size that is centred on the preliminary PLATO NPF field (20~degrees around $\alpha$(2000)~=~265.08003$^\circ$ and $\delta$(2000)~=~39.58370$^\circ$, hereafter the second field). We constructed a colour-magnitude diagram for all selected stars in the second field. The target list consisted of \allnpf~objects, and we observed all of them during the period of 2017--2018. We used the 1.65~m telescope and the high-resolution VUES spectrograph with the full visible light wavelength coverage. The primary data reduction and calibration procedures for VUES data are described in the paper by \citet{Jurgenson2016}. The colour-magnitude diagram of stars from both fields is presented in Fig.~\ref{fig:cmd_targets}.

%  \begin{figure}[htb]

\section{Method of analysis}
\label{sec:methods}

The method of analysis used in this study was presented in \citetalias{Mikolaitis2018}, here we only briefly recall most relevant information for the reader and the new information that is necessary.

Radial velocities were derived the same way as in \citetalias{Mikolaitis2018} using the standard cross-correlation method which also helped to identify the fast-rotating stars and the double-line binaries. Because the target list was formed using the photometric indices, the number of fast-rotating stars or double-line binaries was unknown.

\begin{figure}[]
  \advance\leftskip 0cm
     \centering
   \includegraphics[width=\columnwidth]{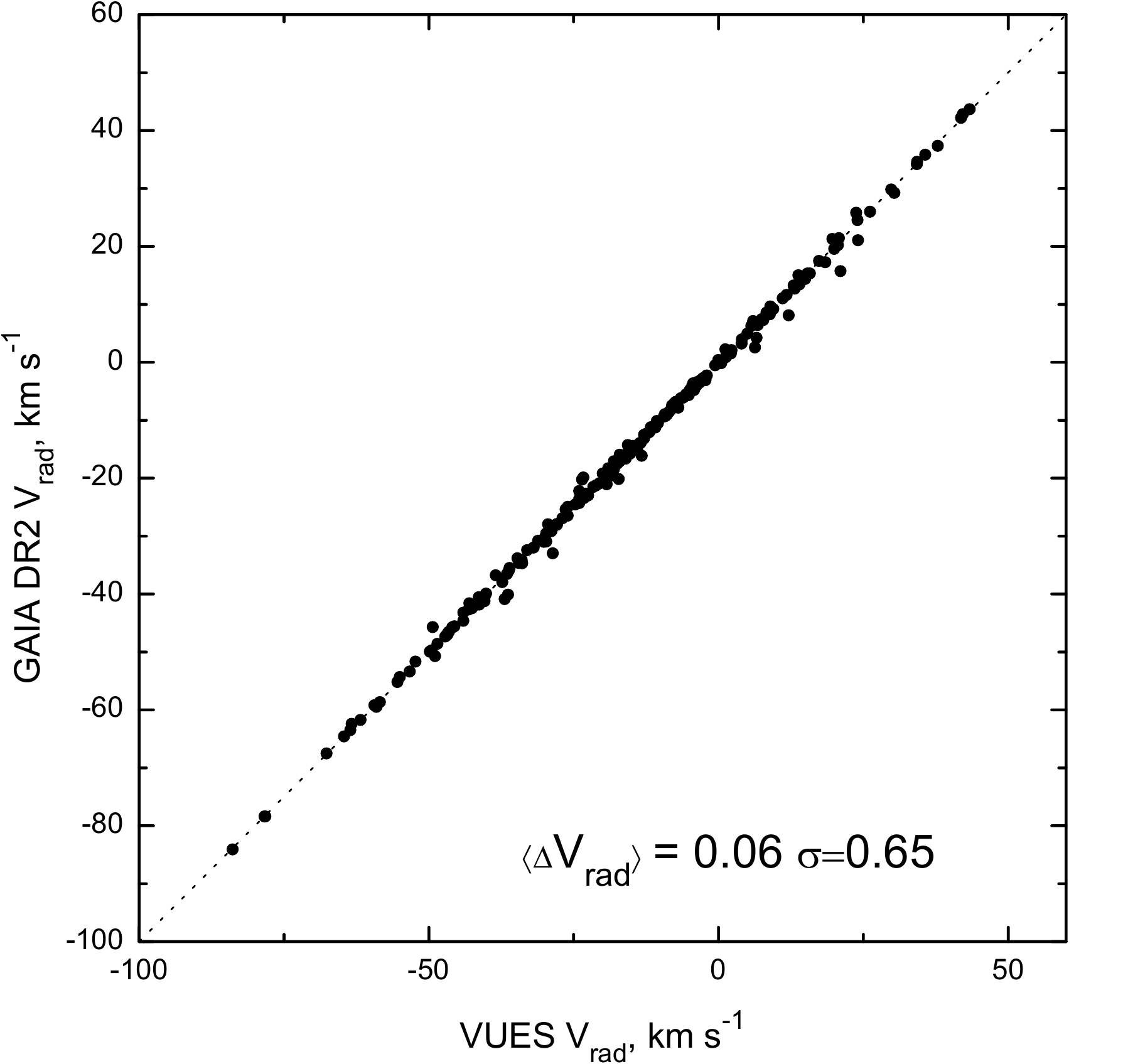}
 \caption{Comparison of radial velocities derived in this study and in {\it Gaia} DR2 (\numrvincommon~stars). The dashed line with a slope of 1 is shown for comparison.}
 \label{fig:vrad_gaia}
 \end{figure}

For the second field we found \numrvincommon~stars within our data set in common with the {\it Gaia} DR2 catalogue that have the radial velocity data. In Fig.~\ref{fig:vrad_gaia} we show a comparison between the {\it Gaia} DR2 radial velocity values with those determined in this study. The mean and standard deviation of differences between the two sets is $\langle \Delta V_{rad} \rangle=0.06\pm 0.65$~km\,s$^{-1}$. 
Due to the broadening and blending of the spectral lines we were not able to measure equivalent widths for \rotatenpf~stars; 
further analysis led us to exclude \binarynpf~stars that display double-line features. Consequently, the final sample of \parametersnpf~stars from the second field was used for the further analysis.

%\subsection{Kinematic properties}
%\label{sec:kinematics}

Kinematic parameters were calculated the same way as in \citetalias{Mikolaitis2018} using the python based package for galactic-dynamics calculations \textit{galpy}\footnote{http://github.com/jobovy/galpy} by \citet{Bovy15}. 
The required input data (parallaxes, proper motions and coordinates) were gathered from the {\it Gaia} DR2 catalogue \citep{Luri2018, Katz2019, Gaia2016, Gaia2018}. We used the radial velocities determined in this work. We note, however, that in \citetalias{Mikolaitis2018} we used the The Tycho-Gaia Astrometric Solution (TGAS) catalogue by \citep{Michalik15}; thus, in this work we decided to recompute the kinematic values for \parametersfull~stars in both fields uniformly in the light of the DR2 release of the {\it Gaia} data.
We integrated the orbits for 5~Gyrs and determined the orbital parameters mentioned earlier. To account for the observational errors, for each object we performed a 1000~MC calculations taking into account the uncertainties in the input parameters (parallaxes, proper motions, and radial velocities).
We used the input values for the position and the movement of the Sun as follows: $R_{\rm gc\odot}=8$~kpc, $V_{0}=220$~km\,s$^{-1}$ \citep{Bovy2012}, the distance from the Galactic plane $z_{0}=0.02$~kpc \citep{Joshi07}, and the local standard of rest (\textit{U, V, W}) = (11.1, 12.24, 7.25)~km\,s$^{-1}$ \citep{Schonrich10}.

Stellar atmospheric parameters (effective temperature $T_{\rm eff}$, surface gravity ${\rm log}~g$, metallicity $\langle{\rm{[Fe/H]}}\rangle$\footnote{We use the notation [Fe/H] for the metallicity delivered using the equivalent width method and [\ion{Fe}{I}/H] for the neutral iron abundance computed using spectral synthesis (see Section~\ref{sec:abundances}).}, and microturbulence velocity $v_{\rm t}$) were determined for \parametersnpf~stars. We used the Vilnius node analysis pipeline that was constantly used for the {\it Gaia}-ESO survey computations (\citealt{Smiljanic2014}). The method employs the DAOSPEC (\citealt{Stetson2008}) software, the MOOG code (\citealt{Sneden1973}), and a grid of MARCS stellar atmosphere models (\citealt{Gustafsson2008}).
Together with the stars from \citetalias{Mikolaitis2018}, \parametersfull~stars from the two fields were selected for the chemical abundance analysis.

\subsection{Stellar ages}
\label{sec:ages}

To calculate the ages of our stars, we used the code UniDAM (the unified tool to estimate distances, ages and masses) by \citet{Mints17}. The code uses a Bayesian approach and the PARSEC isochrones \citep{Bressan12}. As input we use the stellar atmospheric parameters 
% ($T_\mathrm{{eff}}$, log\,\textit{g}, {\bf and} [Fe/H]} 
determined in this work together with the \textit{J}, \textit{H}, and \textit{K} magnitudes from the 2MASS survey \citep{Skrutskie06}.

\subsection{Determination of chemical abundances}
\label{sec:abundances}

Elemental abundances were computed using the spectral synthesis method already described in \citet{Mikolaitis2017} for the following species: \ion{Na}{I}\footnote{For the neutral and ionised element abundances, we use the notations [\ion{El}{I}/H] and [\ion{El}{II}/H].}, \ion{Mg}{I}, \ion{Al}{I}, \ion{Si}{I}, \ion{Si}{II}, \ion{S}{I}, \ion{K}{I}, \ion{Ca}{I}, \ion{Ca}{II}, \ion{Sc}{I}, \ion{Sc}{II}, \ion{Ti}{I}, \ion{Ti}{II}, \ion{V}{I}, \ion{Cr}{I}, \ion{Cr}{II}, \ion{Mn}{I}, \ion{Fe}{I}, \ion{Fe}{II}, \ion{Co}{I}, \ion{Ni}{I}, \ion{Cu}{I}, and  \ion{Zn}{I}.
%only for the \parametersfull~stars with derived atmospheric parameters. 
Atomic lines were selected from the {\it Gaia}-ESO Survey line list  (\citealt{Heiter2015}) and are presented in Table~\ref{tab:TABLE_LINES}. We have also added the following molecular line lists: $\rm{C}_{2}$~\citep[][]{Brooke2013,Ram2014}; CN~\citep[][]{Sneden2014}; CH~\citep[][]{Masseron2014}; SiH~\citep[][]{Kurucz1993}; FeH~\citep[][]{Dulick2003}; CaH~(Plez, Priv. comm.); and OH, MgH, NH~(Masseron, Priv. comm.).
The method employs MARCS (\citealt{Gustafsson2008}) model atmospheres and the spectral synthesis code TURBOSPECTRUM (\citealt{Alvarez1998}).

The algorithm for the abundance determination consists of (I) performing the radial velocity correction, (II) extracting the portion of the line list over the selected wavelength around the spectral line, (III) computing the synthetic spectrum, (IV) adjusting the local continuum, and (V) searching for the best synthetic spectra fit with the observed spectrum applying the minimisation of the $\chi^2$ parameter.
The algorithm works in two steps.
Firstly, it estimates the average line broadening for the star ($v{\rm sin} i$). The $v{\rm sin}i$ is determined by allowing the algorithm to iterate using two free parameters (abundance and $v{\rm sin} i$) and searching of the minimal $\chi^2$ of the fit
% between the synthetic spectrum and observed spectrum 
around 139 neutral iron lines. 
Then the median $v{\rm sin}i$ is fixed and  the abundances were determined by running the algorithm again for all lines of all elements in our line list.

\begin{table}
\caption{Atomic lines ($\lambda, \AA$) used in analysis with the HFS data accounted; the full line list is available online.
}
%\vspace{0.2cm}
\resizebox{\columnwidth}{!}{%
\begin{tabular}{lllll}
\hline\hline
\ion{Mn}{I} \\
4783.4$^B$ & 4823.5$^B$ & 5004.9$^{BW}$ & 5117.9$^J$ & 5255.3$^J$ \\
5394.7$^D$ &5407.4$^{De}$ & 5420.4$^{De}$ & 5432.5$^D$ & 5516.8$^{De}$ \\
 6013.5$^{H}$ & 6016.7$^{H}$ & 6021.8$^{H}$ & 6440.9$^{H}$ &        \\
\ion{Cu}{I} \\
5105.5$^{F}$ & 5218.2$^{He}$ & 5220.1$^{He}$ & 5700.2$^{Be}$ & 5782.1$^{Be}$ \\ 
\hline
 \label{tab:TABLE_LINES}
\end{tabular}%
}
%\begin{tablenotes}
References for the HFS data: $^B$\citet{Brodzinski1987}, $^{BW}$\citet{Blackwell-Whitehead2005a}, $^J$\citet{Johann1981}, $^D$\citet{Davis1971}, $^{De}$\citet{Dembczynski1979}, $^{H}$\citet{Handrich1969}, $^{F}$\citet{Fischer1967}, $^{He}$\citet{Hermann1993}, $^{Be}$\citet{Bergstrom1989}
\\
\end{table}

\subsection{Errors on atmospheric parameters}
\label{sec:erroratmospheres}

The errors on the atmospheric patameters were estimated the same way as in \citetalias{Mikolaitis2018}.

As representative stars for our sample we chose TYC~4573-1916-1 and the Sun, for which we determined the scatter of results while artificially degrading their spectra to the signal-to-noise (S/N) ratios of 25, 50, and 75 per pixel. In this way we were able to estimate the errors caused by the continuum placement and equivalent width measurement.
The precision of measurement was clearly better for the best S/N spectra. The largest errors were found for S/N~=~25 spectra $\sigma_{T_{\rm eff}}$=52~K, $\sigma_{{\rm log}~g}$=0.1~dex, $\sigma_{{\rm [Fe/H]}}$=0.06~dex, $\sigma_{v_{\rm t}}$=0.1~${\rm km~s}^{-1}$ (see Table~3 in \citetalias{Mikolaitis2018} for more details).

The previous example displays the effects of S/N on the atmospheric parameters; however, atmospheric parameters that are based on iron line measurements also suffer from uncertainties in the atomic parameters of the used lines. All these effects create a scatter of measured iron abundances and also an error in linear regression fit that can be directly propagated to follow the uncertainties of atmospheric parameters.
Therefore, the uncertainties for each of the main atmospheric parameters are provided for every star in Table~\ref{tab:CDS} (available online) and they are computed the same way as published in the description of the Vilnius node by \citet{Smiljanic2014}.
The median errors measured by the algorithm in the full \parametersfull~sample (both fields) are 
$\sigma_{T_{\rm eff}}$=46~K, $\sigma_{{\rm log}~g}$=0.3~dex, $\sigma_{{\rm [Fe/H]}}$=0.11~dex, and $\sigma_{v_{\rm t}}$=0.27~${\rm km~s}^{-1}$.

\subsection{Errors on chemical abundances}
\label{sec:errorabundances}

Firstly, we tested the sensitivity of the analysis method to the noise in  $\chi^2$ fitting and continuum placement. For this we used the Monte Carlo simulations in order to add the artificial noise to a statistically significant set of spectra and followed the scatter of abundance values. 
Then, we studied the propagation of errors from the model atmosphere parameters to abundances:
\begin{itemize}
\item 
For the Monte Carlo simulations we used the same representative stars and their degraded spectra as in Section~\ref{sec:erroratmospheres}. We computed 100 abundances for each line at each S/N value. This helped us to estimate the sensitivity of our abundances to the quality of the spectrum. These sensitivities are provided in the form of the standard deviation in Table~\ref{tab:montecarlo}.
\item
The evaluation of the line-to-line scatter is the way to estimate random errors. However, this error estimate is only robust when there are enough lines. We provide these error estimates for all elements in column~6 of Table~\ref{tab:sensitivity}, where $\sigma_{\rm{scat}}^{*}$ is the median of the standard deviation for a given element.
\item
The uncertainties on the main atmospheric parameters that were derived in Section~\ref{sec:erroratmospheres} were propagated into the errors of chemical abundances. The median errors of this type over the sample are provided in Table~\ref{tab:sensitivity}.                                                                                            \end{itemize}

The final error for every element for every star in Table~\ref{tab:CDS} (available online) is a quadratic sum of effects due to uncertainty in four atmospheric parameters and the abundance scatter given by the lines.

\begin{table}
\caption{Errors due to the uncertain continuum placement and line fitting as evaluated by the Monte Carlo simulations.}
  %\begin{threeparttable}
%\vspace{0.2cm}
\centering
\begin{tabular}{lrrr}
%\smallskip
\hline\hline
 & S/N=25 & S/N=50 & S/N=75 \\
\hline
\smallskip
%\smallskip
& \multicolumn{2}{c}{TYC~4573-1916-1} & \\
\multicolumn{4}{c}{$T_{\rm eff}=6153$\,K, ${\rm log}~g=4.01$, ${\rm [Fe/H]}=-0.07$}\\
\smallskip
${\rm[\ion{Na}{I}/H]}$	&	0.09	&	0.08	&	0.04	\\
${\rm[\ion{Mg}{I}/H]}$	&	0.06	&	0.06	&	0.04	\\
${\rm[\ion{Al}{I}/H]}$	&	0.08	&	0.07	&	0.05	\\
${\rm[\ion{Si}{I}/H]}$	&	0.06	&	0.06	&	0.06	\\
${\rm[\ion{Si}{II}/H]}$	&	0.08	&	0.08	&	0.04	\\
${\rm[\ion{S}{I}/H]}$	&	0.10	&	0.09	&	0.08	\\
${\rm[\ion{K}{I}/H]}$	&	0.12	&	0.10	&	0.09	\\
${\rm[\ion{Ca}{I}/H]}$	&	0.09	&	0.09	&	0.06	\\
${\rm[\ion{Ca}{II}/H]}$	&	0.09	&	0.09	&	0.06	\\
${\rm[\ion{Sc}{I}/H]}$	&	0.10	&	0.09	&	0.06	\\
${\rm[\ion{Sc}{II}/H]}$	&	0.11	&	0.09	&	0.07	\\
${\rm[\ion{Ti}{I}/H]}$	&	0.08	&	0.08	&	0.06	\\
${\rm[\ion{Ti}{II}/H]}$	&	0.08	&	0.07	&	0.04	\\
${\rm[\ion{V}{I}/H]}$	&	0.07	&	0.07	&	0.05	\\
${\rm[\ion{Cr}{I}/H]}$	&	0.08	&	0.06	&	0.05	\\
${\rm[\ion{Cr}{II}/H]}$	&	0.09	&	0.07	&	0.05	\\
${\rm[\ion{Mn}{I}/H]}$	&	0.10	&	0.08	&	0.05	\\
${\rm[\ion{Fe}{I}/H]}$	&	0.08	&	0.06	&	0.04	\\
${\rm[\ion{Fe}{II}/H]}$	&	0.10	&	0.09	&	0.06	\\
${\rm[\ion{Co}{I}/H]}$	&	0.08	&	0.07	&	0.02	\\
${\rm[\ion{Ni}{I}/H]}$	&	0.07	&	0.06	&	0.03	\\
${\rm[\ion{Cu}{I}/H]}$	&	0.08	&	0.08	&	0.06	\\
${\rm[\ion{Zn}{I}/H]}$	&	0.11	&	0.09	&	0.05	\\
\smallskip
%\hline
& \multicolumn{2}{c}{Sun$^*$} & \\
\multicolumn{4}{c}{$T_{\rm eff}=5779$\,K, ${\rm log}~g=4.49$, ${\rm [Fe/H]}=-0.03$}\\
${\rm[\ion{Na}{I}/H]}$	&	0.07	&	0.06	&	0.04	\\
${\rm[\ion{Mg}{I}/H]}$	&	0.05	&	0.04	&	0.03	\\
${\rm[\ion{Al}{I}/H]}$	&	0.05	&	0.03	&	0.01	\\
${\rm[\ion{Si}{I}/H]}$	&	0.06	&	0.05	&	0.02	\\
${\rm[\ion{Si}{II}/H]}$	&	0.07	&	0.04	&	0.03	\\
${\rm[\ion{S}{I}/H]}$	&	0.10	&	0.07	&	0.07	\\
${\rm[\ion{K}{I}/H]}$	&	0.10	&	0.09	&	0.08	\\
${\rm[\ion{Ca}{I}/H]}$	&	0.07	&	0.04	&	0.03	\\
${\rm[\ion{Ca}{II}/H]}$	&	0.07	&	0.06	&	0.05	\\
${\rm[\ion{Sc}{I}/H]}$	&	0.09	&	0.07	&	0.05	\\
${\rm[\ion{Sc}{II}/H]}$	&	0.11	&	0.10	&	0.07	\\
${\rm[\ion{Ti}{I}/H]}$	&	0.07	&	0.07	&	0.05	\\
${\rm[\ion{Ti}{II}/H]}$	&	0.08	&	0.08	&	0.05	\\
${\rm[\ion{V}{I}/H]}$	&	0.06	&	0.03	&	0.01	\\
${\rm[\ion{Cr}{I}/H]}$	&	0.05	&	0.03	&	0.02	\\
${\rm[\ion{Cr}{II}/H]}$	&	0.08	&	0.07	&	0.07	\\
${\rm[\ion{Mn}{I}/H]}$	&	0.06	&	0.03	&	0.03	\\
${\rm[\ion{Fe}{I}/H]}$	&	0.05	&	0.03	&	0.02	\\
${\rm[\ion{Fe}{II}/H]}$	&	0.09	&	0.08	&	0.06	\\
${\rm[\ion{Co}{I}/H]}$	&	0.08	&	0.07	&	0.05	\\
${\rm[\ion{Ni}{I}/H]}$	&	0.05	&	0.05	&	0.02	\\
${\rm[\ion{Cu}{I}/H]}$	&	0.07	&	0.05	&	0.04	\\
${\rm[\ion{Zn}{I}/H]}$	&	0.08	&	0.07	&	0.04	\\
\hline
\\
\end{tabular}
%\begin{tablenotes}

$^*$Solar atmospheric parameters derived with our method. 
%\end{tablenotes}

%\end{threeparttable}
 \label{tab:montecarlo}
\end{table}

   \begin{table*}
%\begin{center}
\caption{Median effects on the derived [El/Fe] ratios resulting from the atmospheric parameter uncertainties for the sample stars.
}
\label{tab:sensitivity}
      \[
   %%resize%   \resizebox{\columnwidth}{!}{%
         \begin{tabular}{lcrrcrrrr}
            \hline
%	    \hline
            \noalign{\smallskip}
	    El & 
	    ${ \Delta T_{\rm eff} }$ & 
            ${ \Delta \log g }$ & 
            $\Delta {\rm [Fe/H]}$ & 
            ${ \Delta v_{\rm t} }$ 
            & $\sigma_{\rm{scat}}^{1}$  
     	     & $N_{max}^{2}$ 
     	     & $ \sigma_{\rm total\left[\frac{X}{Fe}\right]}^{3} $  
	     & $ \sigma_{\rm all\left[\frac{X}{Fe}\right]}^{4} $  
	     \\ 
	     & K & & & km\,s$^{-1}$ & & & & \\
            \noalign{\smallskip}
            \hline
            \noalign{\smallskip}

Na\,{\sc i} 	&	0.01	&	0.04	&	0.02	&		0.02	&	0.03	&	4	&	0.05	&			0.06	\\
Mg\,{\sc i} 	&	0.01	&	0.06	&	0.02	&		0.03	&	0.07	&	5	&	0.07	&			0.11	\\
Al\,{\sc i} 	&	0.01	&	0.02	&	0.02	&		0.03	&	0.04	&	5	&	0.03	&			0.06	\\
Si\,{\sc i} 	&	0.01	&	0.01	&	0.02	&		0.03	&	0.03	&	14	&	0.03	&			0.04	\\
Si\,{\sc ii} 	&	0.02	&	0.05	&	0.02	&		0.03	&	0.07	&	7	&	0.06	&			0.10	\\
S\,{\sc i} 	&	0.04	&	0.08	&	0.02	&		0.05	&	0.09	&	4	&	0.10	&			0.14	\\
K\,{\sc i} 	&	0.03	&	0.07	&	0.03	&		0.04	&	0.06	&	1	&	0.10	&			0.13	\\
Ca\,{\sc i} 	&	0.02	&	0.07	&	0.02	&		0.03	&	0.05	&	31	&	0.08	&			0.09	\\
Ca\,{\sc ii} 	&	0.02	&	0.06	&	0.03	&		0.04	&	0.06	&	7	&	0.08	&			0.10	\\
Sc\,{\sc i} 	&	0.04	&	0.03	&	0.02	&		0.03	&	0.08	&	7	&	0.06	&			0.11	\\
Sc\,{\sc ii} 	&	0.01	&	0.09	&	0.02	&		0.04	&	0.04	&	12	&	0.10	&			0.11	\\
Ti\,{\sc i} 	&	0.04	&	0.04	&	0.02	&		0.03	&	0.04	&	81	&	0.07	&			0.08	\\
Ti\,{\sc ii} 	&	0.02	&	0.08	&	0.03	&		0.04	&	0.04	&	19	&	0.09	&			0.10	\\
V\,{\sc i} 	&	0.03	&	0.02	&	0.02	&		0.03	&	0.05	&	8	&	0.04	&			0.06	\\
Cr\,{\sc i} 	&	0.02	&	0.03	&	0.02	&		0.03	&	0.04	&	21	&	0.04	&			0.06	\\
Cr\,{\sc ii} 	&	0.02	&	0.08	&	0.02	&		0.04	&	0.04	&	2	&	0.09	&			0.11	\\
Mn\,{\sc i} 	&	0.03	&	0.03	&	0.03	&		0.03	&	0.05	&	14	&	0.06	&			0.08	\\
Fe\,{\sc i} 	&	0.02	&	0.03	&	0.02	&		0.03	&	0.05	&	138	&	0.05	&			0.05	\\
Fe\,{\sc ii} 	&	0.02	&	0.09	&	0.02	&		0.05	&	0.05	&	11	&	0.10	&			0.11	\\
Co\,{\sc i} 	&	0.02	&	0.01	&	0.02	&		0.03	&	0.05	&	7	&	0.03	&			0.06	\\
Ni\,{\sc i} 	&	0.02	&	0.02	&	0.02	&		0.03	&	0.03	&	30	&	0.04	&			0.06	\\
Cu\,{\sc i} 	&	0.03	&	0.02	&	0.03	&		0.03	&	0.05	&	6	&	0.05	&			0.07	\\
Zn\,{\sc i} 	&	0.01	&	0.02	&	0.02	&		0.03	&	0.09	&	3	&	0.04	&			0.10	\\
\hline
         \end{tabular} %%resize%}
      \]

%\begin{tablenotes}
$^{1} \sigma_{\rm{scat}}$ stands for median line-to-line scatter;\\ 
$^{2}{\rm N}_{max}$ presents the number of lines investigated;\\
$^{3}\sigma_{\rm total([X/Fe])} $ stands for the median of the quadratic sum  of all four effects on [El/Fe] ratios; \\
$^{4}\sigma_{\rm all([X/Fe])} $ is the median combined effect of $ \sigma_{\rm total([El/Fe])} $ and the line-to-line scatter $\sigma_{\rm{scat}}$. 
%\end{tablenotes}

   \end{table*}
   
   %%%%%%%%%%%%%%%%%%%%%%%%%%%%%%%%%%%%%%%%%%%%%%%%%%%%%%%%%%%%%%%%%%%%%%%%%%%%%%5

   %%%%%%%%%%%%%%%%%%%%%%%%%%%%%%%%%%%%%%%%%%%%%%%%%%%%%%%%%%%%%%%%%%%%%%%%%%%%%%%%%

\subsection{Testing the validity of the stellar parameters}
\label{sec:ionisation}

\begin{figure*}[htb]
  \centering
   \advance\leftskip 0cm
   \includegraphics[scale=0.65]{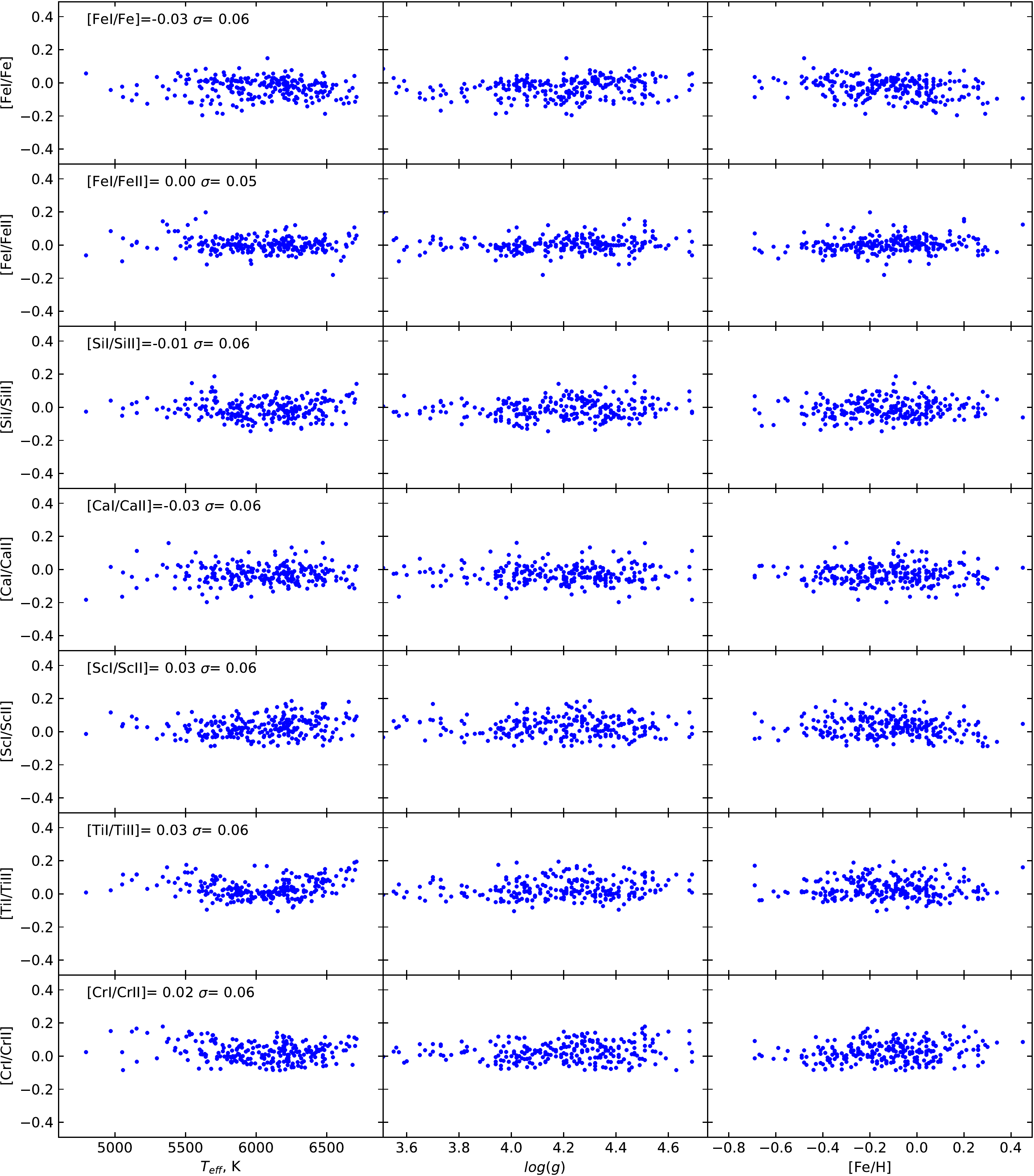}
  \caption{
Differences between the mean abundances of Si, Ca, Ti, and Cr and their ionised counterparts as a function of the atmospheric parameters. 
  }
  \label{fig:Efig:ELEMENTS_ion}
  \end{figure*}

The metallicity value [Fe/H] representing the ionised and neutral iron abundance was derived as the mean metallicity of a star using the equivalent width method. However, we also derived \ion{Fe}{I} and \ion{Fe}{II} using spectral synthesis method. Since the iron abundances were derived using different and slightly independent ways it is important to check if they are consistent. We provide this comparison in the first row of Fig.~\ref{fig:Efig:ELEMENTS_ion}. The bias of [\ion{Fe}{I}/Fe] is +0.03~dex with a scatter $\sigma$=0.06~dex. We did not find any noticeable systematic effects with respect to $T_{\rm eff}$, ${\rm log}~g$, or ${\rm [Fe/H]}$. 

The comparison of neutral and ionised elemental abundances are used to check the ionisation equilibrium. In Fig.~\ref{fig:Efig:ELEMENTS_ion} we provide similar comparisons for [\ion{Fe}{I}/\ion{Fe}{II}], [\ion{Si}{I}/\ion{Si}{II}], [\ion{Ca}{I}/\ion{Ca}{II}], [\ion{Sc}{I}/\ion{Sc}{II}], [\ion{Ti}{I}/\ion{Ti}{II}], and [\ion{Cr}{I}/\ion{Cr}{II}]. The bias is mostly quite small, from $-0.03$~dex for [\ion{Ca}{I}/\ion{Ca}{II}] ($\sigma$=0.06~dex) to 0.04~dex for [\ion{Ti}{I}/\ion{Ti}{II}] ($\sigma$=0.06~dex), whereas the largest scatter is $\sigma$=0.07~dex for [\ion{Sc}{I}/\ion{Sc}{II}]. We did not find noticeable systematic effects with respect to ${\rm log}~g$ or ${\rm [Fe/H]}$. However, [\ion{Ti}{I}/\ion{Ti}{II}] and [\ion{Sc}{I}/\ion{Sc}{II}] versus $T_{\rm eff}$ distributions show some small dependences. Therefore, we conclude that the noticed biases, scatter, and dependences are relatively small and should not affect our study.

%%%%%%%%%%%%%%%%%%%%%%%%%%%%%%%%%%%%%

\subsection{NLTE effects on abundances}
\label{sec:nlte}

There is evidence that silicon and titanium can be affected by non-local thermodynamic equilibrium (NLTE) effects (e.g. \citealt{Bergemann2011, Bergemann2013}). The NLTE effects for silicon abundances are mostly noticeable in supergiant stars (see \citealt{Bergemann2013}); however, \citet{Bergemann2011} and \citet{Zhao2016} recommend taking NLTE into account for lines of ionised titanium in spectra of giant stars as well. Thus, as the NLTE effects for the Ti lines are largest in metal-poor and giant stars \citep{Zhao2016}, it should be safe to use the classical LTE approach to compute $\alpha$-element abundances for our sample stars, which are mostly metal-rich dwarfs. According to the results by \citet{Zhao2016}, Ca and Sc should only have larger NLTE-LTE departures in the low-metallicity regime ([Fe/H]$<-1.5$~dex), which should not impact our sample as well. Magnesium abundances also have very low NLTE departures for the majority of the lines that we used (\citealt{Zhao2016,Alexeeva2018}).

\citet{Zhao2016} showed that sodium abundances are notably sensitive to NLTE effects for metallicities lower than $\approx-1.4$~dex and are almost insignificant for metal-rich dwarfs.
The aluminium abundances can be affected by NLTE (\citealt{Baumueller1997, Gehren2004, Gehren2006, Andrievsky2008}).
\citet{Baumueller1997} calculated the LTE and NLTE aluminium abundances using the spectral lines at $\sim$8772~$\AA$, which we used in our analysis as well. For stars that are similar to the ones analysed in this work, the NLTE abundances are up to 0.07~dex higher than the LTE values, thus the corrections would be similar to the errors arising from the uncertainties in parameters and are quite small.
\citet{Zhao2016} computed the NLTE and LTE abundances for aluminium lines at 6696~$\AA$ and 6698~$\AA$ as well, and showed that there are differences of up to $\approx$0.1~dex. Considering that the differences are quite small, the NLTE-LTE departures should not be the main source of dispersion for Al in this metallicity range.

\citet{Korotin2017} calculated the NLTE corrections for sulphur lines used by the {\it Gaia}-ESO survey and showed that they generally do not exceed 0.07~dex for dwarfs. It was shown by \citet{Zhao2016} and \citet{Takeda2002}, who computed NLTE and LTE abundances for the most prominent K\,{\sc i} 7698~$\AA$ line for F and G dwarfs, that potassium has much more noticeable NLTE effects that range from approximately $-0.3$ to $-0.6$~dex in the metallicity regime, which is similar to the values in our sample. In this study we applied the NLTE corrections from \citet{Takeda2002,Takeda2009}\footnote{http://www2.nao.ac.jp/$\sim$takedayi/potassium$\_$nonlte} for the 7698~$\AA$ potassium line and report the NLTE abundance results.

We note that there should be some NLTE effects influencing the determined manganese abundances. Some NLTE deviations have already been studied in previous works. \citet{Mishenina2015}, who studied stars similar to those in our sample, calculated the average LTE-NLTE abundance variations of $0.01\pm0.04$ and $0.02\pm0.04$~dex for the thin- and thick-disc stars, respectively. \citet{Bergemann2008} also showed that the NLTE corrections for manganese in metal-rich dwarfs should be around 0.05~dex. \citet{Battistini2015} reported the NLTE abundance  corrections of $0.059\pm0.046$~dex for stars similar to ours (${\rm [Fe/H]}>-0.7$~dex). They reported the $0.069\pm0.036$~dex NLTE abundance corrections for cobalt as well. Vanadium, chromium,  and nickel were barely studied using the NLTE approach. The NLTE correction for the solar abundances of Ni and Cr are negligible, whereas the corrections for vanadium could reach $\approx$0.1~dex (see \citealt{Scott2015,Sneden2016}). Therefore, the classical LTE determinations for the iron-peak element abundances (V, Cr, Co, Ni, Mn) can be accepted within the given uncertainty of 0.1~dex.

The NLTE effects for the Cu abundances should not be significant to our results. \citet{Shi2014} and \citet{Yan2016} have shown that the NLTE-LTE corrections for Cu abundances in stars similar to the values in our sample are not very large (up to 0.1~dex for [Fe/H]$\approx-1.0$). Moreover, \citet{Zhao2016} have shown that the Cu NLTE effects are not very significant in a star sample similar to ours. The \ion{Zn}{I} 6362~$\AA$ line is affected by the broad auto-ionisation absorption feature of \ion{Ca}{I} as identified by \citet{Mitchell1965}. \citet{Chen2004} showed that diferentially with respect to the Sun, the largest effect of the mentioned \ion{Ca}{I} feature would only be about 0.02~dex. That is a significantly smaller value than our typical error of Zn abundance, thus it has been neglected in the present analysis.

%%%%%%%%%%%%%%%%%%%%%%%%%%%%%%%%%%%%%

\section{Atmospheric parameters}
\label{sec:atmospheres}

\begin{figure*}[!htb]
   \advance\leftskip 0cm
      \centering
   \includegraphics[scale=1.5]{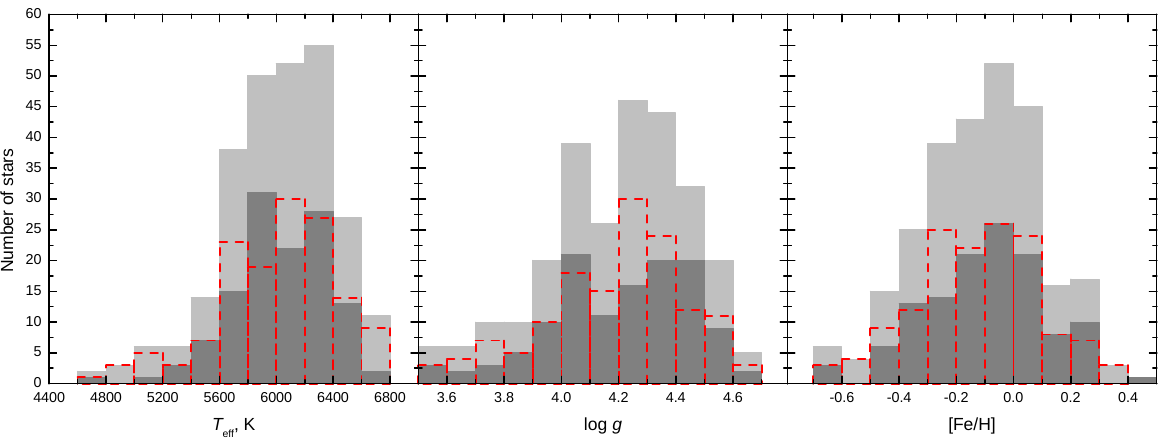}
  \caption{Histograms of the determined spectroscopic parameters ($T_{\rm eff}$, log\,$g$,  and [Fe/H]) for all stars in our sample, the dark grey columns are for the NPF field, the red dashed lines are for the STEP~02 field, and the light grey columns are for both fields together.}
  \label{fig:resultsHISTOGRAM}
  \end{figure*}

\begin{figure}[!htb]
   \advance\leftskip 0cm
      \centering
    \includegraphics[width=\columnwidth]{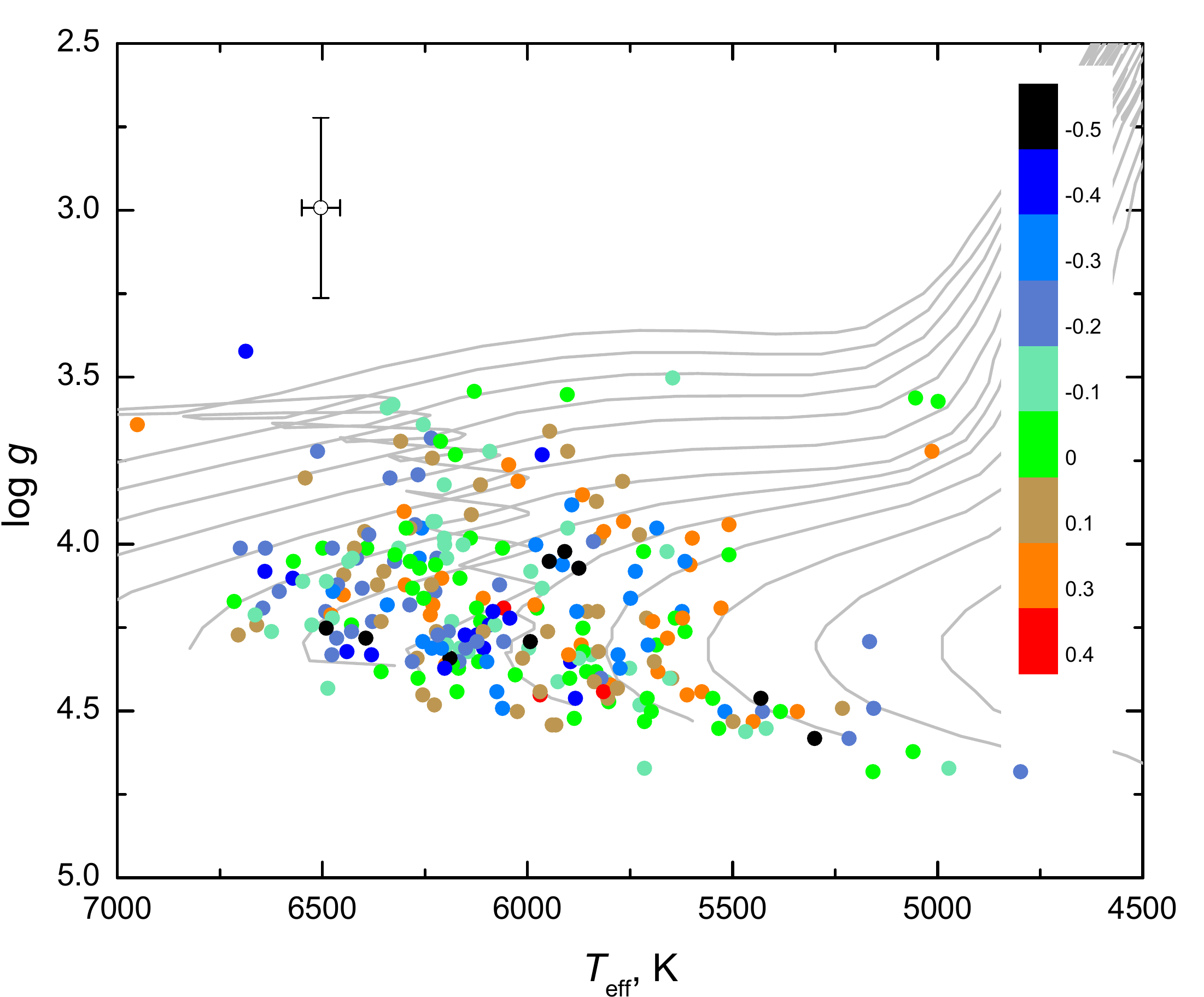}
  \caption{Temperature--gravity diagram of \parametersfull~stars (dots) of the investigated fields with colour-coded metallicity. Evolutionary tracks with masses between 0.7 and 1.9~$M_{\odot}$ and $Z_{\rm ini}$=0.019 by \citet{Girardi2000} are plotted as grey solid lines. }
  \label{fig:hr}
\end{figure}

In this study, we determined atmospheric parameters for \parametersnpf~stars of the NPF field to complement the previously determined \parametersstep~results of the STEP~02 field. The stellar parameters are provided in Table~\ref{tab:CDS}. Distributions of the determined atmospheric parameters for the stars in both fields are shown in Fig.~\ref{fig:resultsHISTOGRAM}. They are quite similar: the effective temperatures of the stars range from $T_{\rm eff}\approx$4700~K to 6900~K (Fig.~\ref{fig:resultsHISTOGRAM}a) with a peak at $\approx$6000~K; ${\rm log}\,g$ values range from $\approx$3.5 to 4.7 (Fig.~\ref{fig:resultsHISTOGRAM}b); and [Fe/H] values range from $\approx-0.7$ to +0.4~dex (Fig.~\ref{fig:resultsHISTOGRAM}), where the majority of the stars have approximately solar [Fe/H]. In addition, we plot the ($T_{\rm eff}$, ${\rm log}\,g$) diagram with colour-coded metallicity in Fig.~\ref{fig:hr}. The stellar evolutionary tracks by \citealt{Girardi2000} are plotted in the background. Their masses are between 0.7 and 1.9~$M_{\odot}$ and the initial metallicity $Z_{\rm ini}$=0.019.

\begin{figure}[htb]
 \advance\leftskip 0cm
 \centering
  \includegraphics[width=\columnwidth]{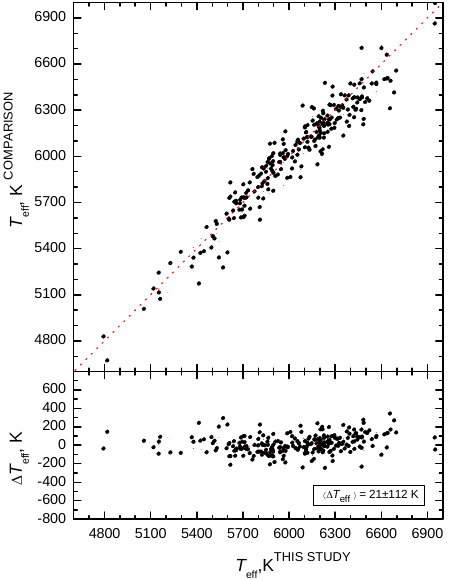}
  \caption{Comparison between $T_{\rm eff}$ values of our study with values from the {\it Gaia}~DR2 (245 stars, black dots). The red dashed lines with a slope of 1 are shown for comparison.}
  \label{fig:resultsCOMAPRISON}
\end{figure}

In \citetalias{Mikolaitis2018} we compared our results with other studies. The comparison with a number of results from other spectroscopic studies (27\% of stars in common) showed rather good compatibility ($\langle \Delta T_{\rm eff} \rangle$=7$\pm$73~K, $\langle \Delta {\rm log}~g \rangle=0.02\pm$0.09, and $\langle \Delta {\rm [Fe/H]} \rangle=-0.02\pm0.08$~dex). The scatter of our results was slightly higher ($\langle \Delta T_{\rm eff} \rangle$=14$\pm$125~K, $\langle
\Delta {\rm log}~g \rangle=0.02\pm$0.16, and $\langle \Delta {\rm [Fe/H]}
\rangle=-0.02\pm0.09$~dex) with respect to  the values from the photometric study by \citet{Casagrande2011}. Additionally, we show a comparison of our effective temperatures (both fields) with the {\it Gaia}~DR2 results (245 stars in common) in Fig.~\ref{fig:resultsCOMAPRISON}.

\section{Kinematic parameters and ages}
\label{sec:kinematicparameters}

\begin{figure*}[!htb]
   \advance\leftskip 0cm
      \centering
   \includegraphics[scale=0.60]{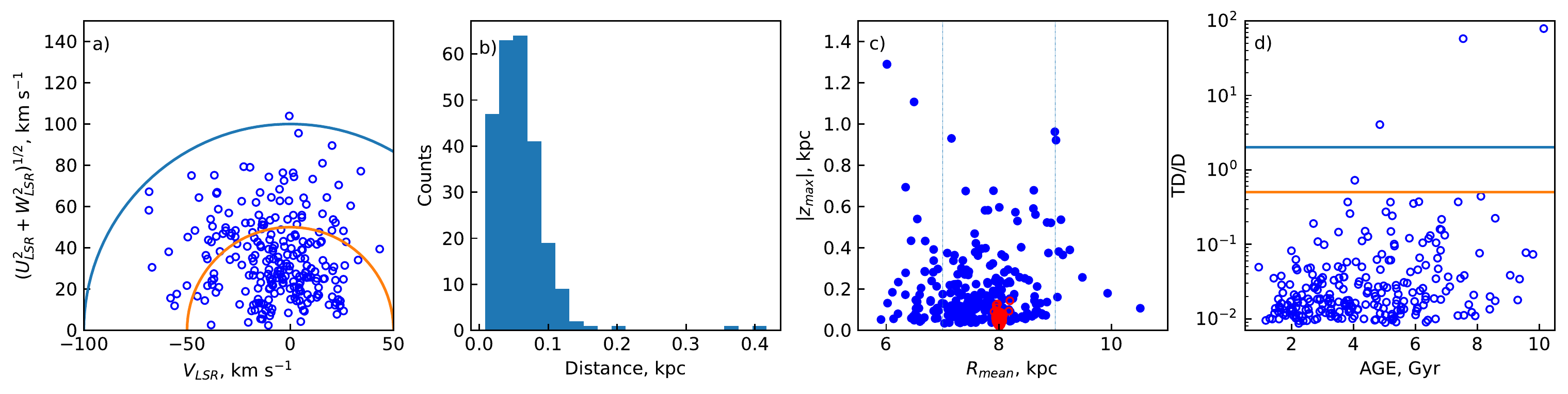}
  \caption{Kinematic parameters: a) Toomre diagram of sample data with lines that show constant values of the total space velocity ($v_{\rm tot}=(U_{\rm LSR}+V_{\rm LSR}+W_{\rm LSR})^{\rm 1/2)}$) at 50 and 100 km\,s$^{\rm -1}$; b) histogram of distances of sample stars; c) distribution of sample stars in $z_{\rm max}$ vs. $R_{\rm mean}$ plane, where the two vertical dashed lines delimit the solar neighbourhood 7$<R_{\rm gc}<$9~kpc and red circles are current positions ($|z|$ vs. $R$) for comparison; d) kinematical thick-to-thin disc probability ratios (TD/D) vs. age for the sample stars, where the upper and lower lines mark TD/D=2.0 and TD/D=0.5, respectively.}
  \label{fig:resultsKINEMATICS}
  \end{figure*}

As we note in Sec.~\ref{sec:methods} of this work, we determined stellar kinematic parameters for both fields using the new {\it Gaia} DR2 data. In Fig.~\ref{fig:resultsKINEMATICS} we present the kinematic distribution of our sample stars (\parametersfull~objects). 

A Toomre diagram is a way to understand the sample in the context of Galactic populations. It helps in the study of stellar combined vertical and radial kinetic energies versus the rotational energy (\citealt{Bensby2014}). The stars with a lower combined velocity $v_{\rm tot}<50$~km~s$^{\rm -1}$ ($v_{\rm tot}=(U_{\rm LSR}+V_{\rm LSR}+W_{\rm LSR})^{\rm 1/2)}$) are most probably the thin-disc star, whereas those with $v_{\rm tot}>50$~km~s$^{-1}$ should belong to the thick-disc (see Fig.~\ref{fig:resultsKINEMATICS} panel a). 

Another way to disentangle the sample is to study the thick-to-thin disc probability ratios (\textit{TD/D}). The method that we used to compute the \textit{TD/D} is described in \citet{Bensby2003, Bensby2005, Bensby2014}, where it was advised to classify a star as thick-disc by requiring it to have a probability at least two times larger than that of being a thin-disc star. Thus, all stars with \textit{TD/D}$>$2.0 should be the thick-disc candidates and \textit{TD/D}$<$0.5 should be thin-disc candidates. 

Following the first criterion we selected 172 thin-disc and 25 thick-disc stars, and 57 should be transition stars between the thin- and thick-discs. The second criterion indicated only three thick-disc stars and only one star would be intermediate.

\begin{figure}[htb]
   \advance\leftskip 0cm
   \centering
    \includegraphics[width=\columnwidth]{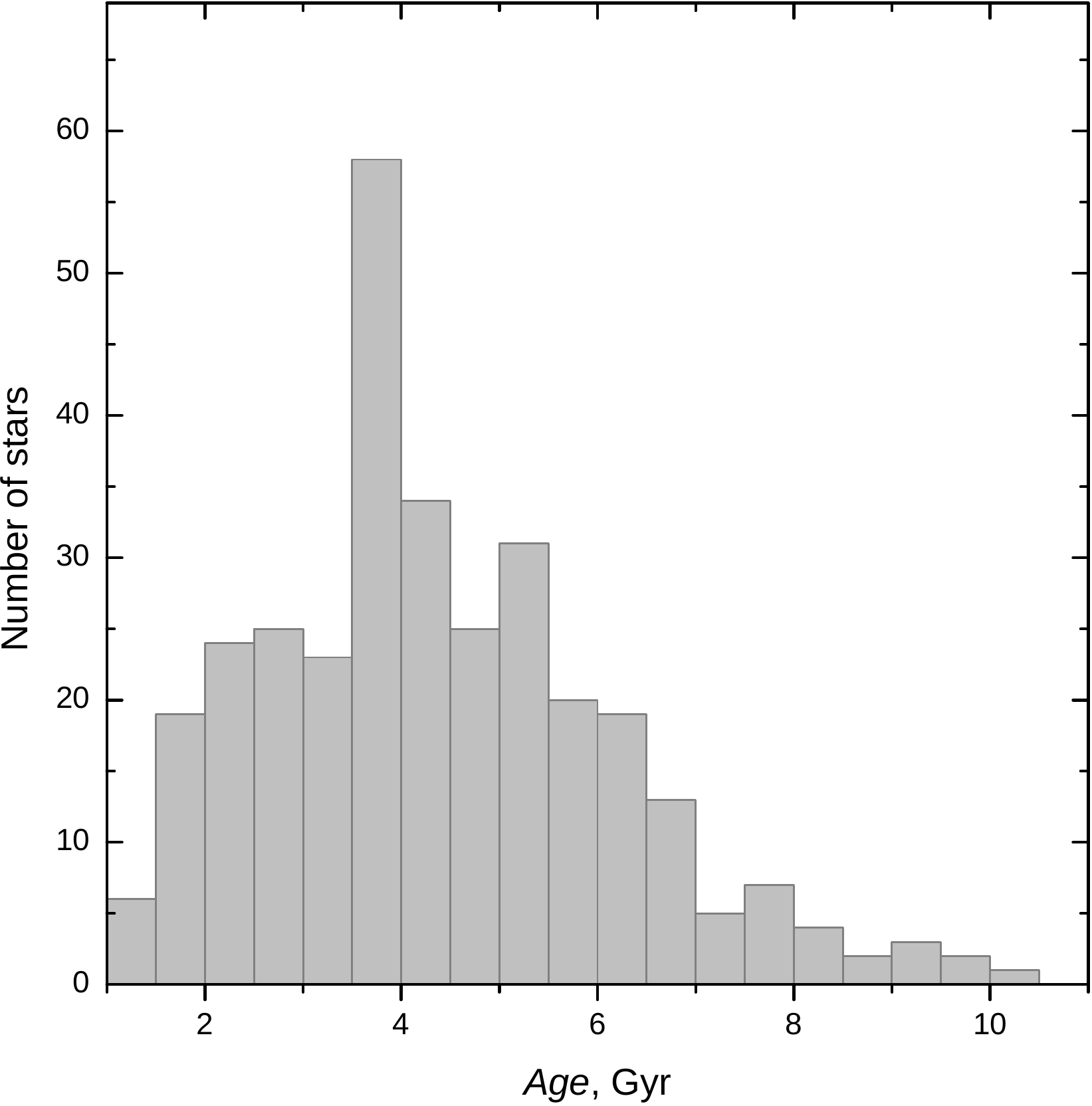}
  \caption{Age histogram of the sample stars.}
  \label{fig:age_histogram}
  \end{figure}

As is seen from Fig.~\ref{fig:age_histogram}, the ages of the stars in our sample range from 1 to 10~Gyr; the majority have ages close to solar, about 4~Gyrs.
\citet{Haywood2013} suggest that stars younger than $\approx$8~Gyr should be from the thin-disc population and those older than $\approx$8~Gyr are either from the thick-disc population or the metal-poor thin-disc. Our sample is composed of mostly younger generation stars: 238 stars are younger than $\approx$8~Gyr and only 12 stars are $\approx$8~Gyr to $\approx$10~Gyr old. However, we have only two stars that have probability ratio \textit{TD/D} close to 100~\% and both are found in the older part of panel d in Fig.~\ref{fig:resultsKINEMATICS}. 

The chemical division of thin- and thick-disc populations that is based on $\alpha$-to-iron abundance ratios can be also useful. We used this method in our work, as discussed in the next section.

  \begin{figure}[htb]
  \centering
   \advance\leftskip 0cm
    \includegraphics[width=\columnwidth]{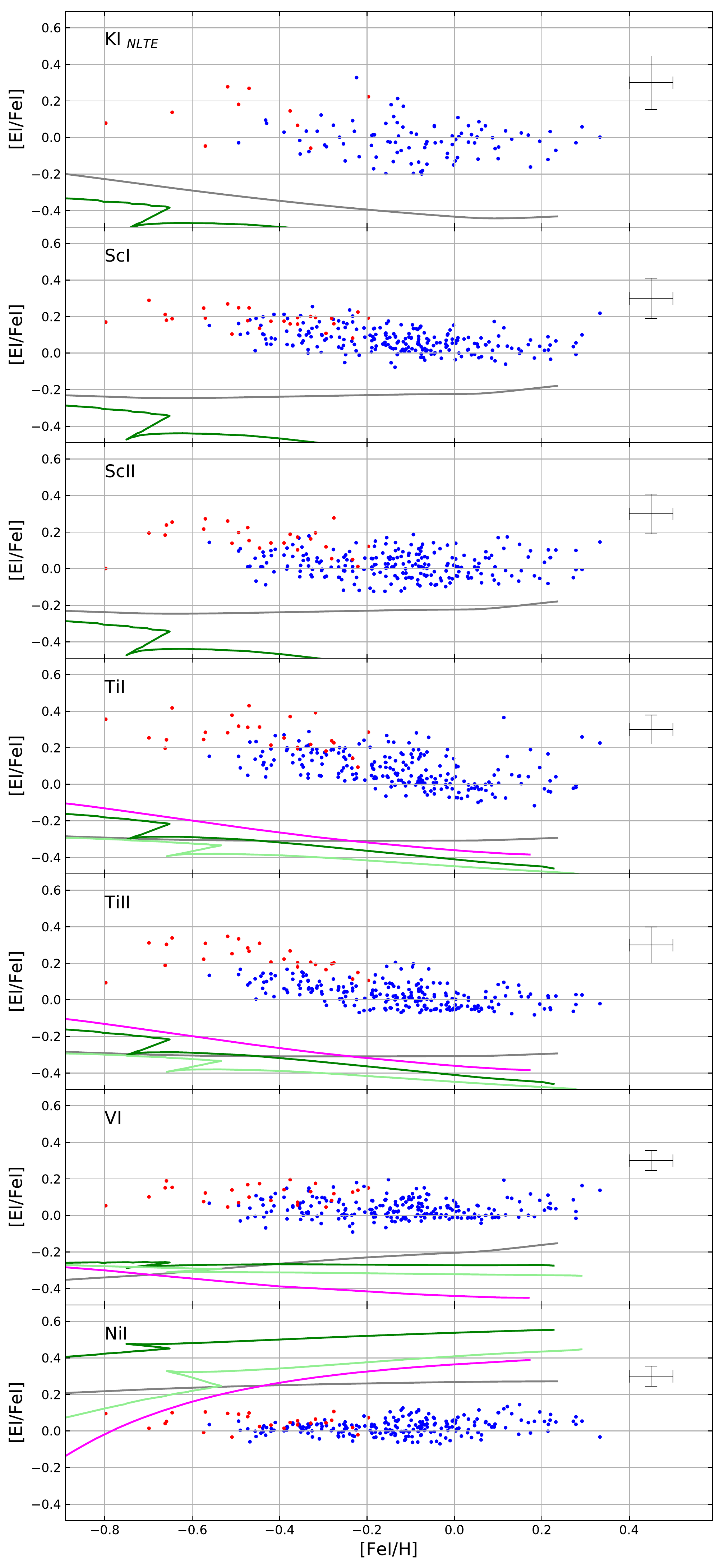}
  \caption{
Observed element-to-iron abundance ratios as a function of the [\ion{Fe}{I}/H] abundance. Cases where all models fail to represent the data. Shown are the \citet{Romano2010} R1 model (dark green line) and R15 model (light green line), the \citet{Kobayashi2011} thin disc model (magenta line), and the \citet{Prantzos2018} baseline model (grey line).
  }
  \label{fig:ELEMENTS_separation_failed}
  \end{figure}

  \begin{figure}[htb]
  \centering
   \advance\leftskip 0cm
      \includegraphics[width=\columnwidth]{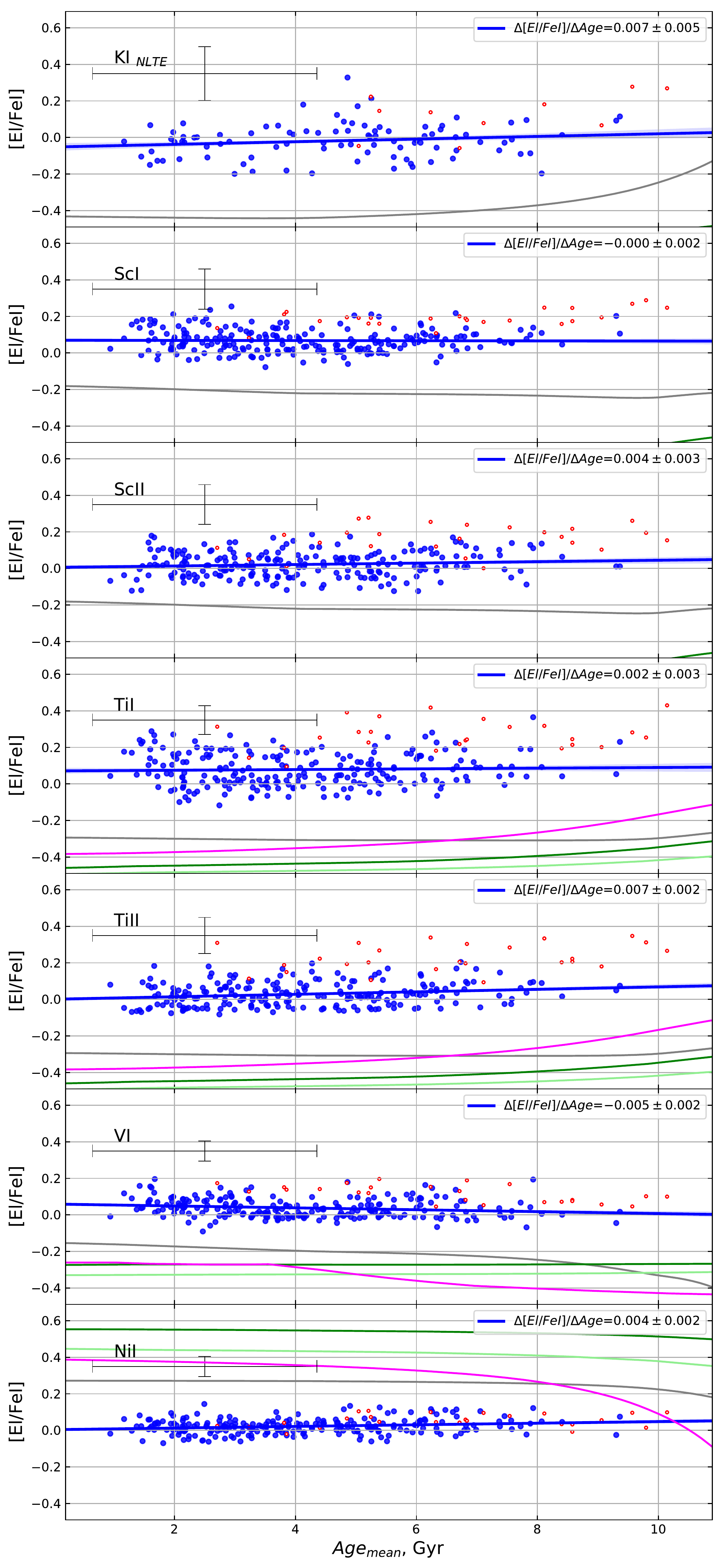}
  \caption{
Observed element-to-iron abundance ratios as a function of age for the cases where all models fail to represent the data. The blue line shows the $\Delta {\rm [El/Fe I]}/\Delta {\rm Age}$ gradient. All other notations are as in Fig.~\ref{fig:ELEMENTS_separation_failed}.
  }
  \label{fig:ELEMENTS_age_failed}
  \end{figure}

\section{Elemental abundance ratios and comparison with models}
\label{sec:ratios}

It is important to study the chemical abundance ratios and their patterns in spatial and temporal context as they are fingerprints of the Galactic chemical evolution history. In Figures~\ref{fig:ELEMENTS_separation_failed}~and~\ref{fig:ELEMENTS_separation} we show the ratios [El/\ion{Fe}{I}] versus [\ion{Fe}{I}/H]. This way of representing elemental abundance ratios is commonly used to chemically disentangle Galactic substructures (thin-disc, thick-disc, and halo) which have different chemical properties (e.g. \citealt{Fuhrmann2004,Neves2009,Bensby2003,Bensby2014,Adibekyan2012,Mikolaitis2017}). This so-called chemical tagging is one of the ways to separate thin- and thick-discs.
In particular, it has been shown that the chemical distinction of the thin- and thick-discs is possible by using [$\alpha$/Fe] ratios and it has been used in a number of studies (e.g. \citet{Fuhrmann1998, Recio2014, Rojas2016, Mikolaitis2014, Mikolaitis2017}). In \citet{Mikolaitis2014, Mikolaitis2017} we already discussed that [Mg/Fe] is one of the best indicators to define the Galactic disc components. We used the same method as in \citet{Mikolaitis2014, Mikolaitis2017} to chemically tag \numthin~stars as thin ($\alpha$-rich) disc members and \numthick~stars as thick ($\alpha$-poor) disc members in our sample. In the [\ion{Mg}{I}/\ion{Fe}{I}] versus [\ion{Fe}{I}/H] plot in Fig.~\ref{fig:ELEMENTS_separation} we display both chemically defined populations in different colours: blue for the thin-disc and red for the thick-disc stars. We keep this colouring style throughout the entire paper for the two discs. We also show the chemical abundance ratio variations in relation to the age, $R_{\rm{mean}}$, $|z_{\rm{max}}|$ in Figures~\ref{fig:ELEMENTS_age_failed},~\ref{fig:ELEMENTS_age},~\ref{fig:ELEMENTS_rmean},~and~\ref{fig:ELEMENTS_zmax}, where we provide the numerical evaluation of the gradients as well. 

Moreover, chemical abundance ratios can be used as a test for the Galactic chemical evolution (GCE) models. A good GCE model would have to explain the dynamical properties of the disc, and also its chemical element distributions according to metallicity and age or radial and vertical gradients.
The GCE models differ in their internal formalism, input parameters, and various assumptions (see \citealt{Nomoto2013}). For example, the different input parameter combinations that can be used in GCE models can include different types of supernovae, different chemical element yields which can be used to constrain the nucleosynthesis scenarios, and nucleosynthesis timescales.

Our study is based on the sample of stars that are very well confined in the solar vicinity. As can be seen from the second panel in Fig.~\ref{fig:resultsKINEMATICS}, all stars are less than 0.42~kpc from the Sun, and most are closer than 0.15~kpc. We compare several recent GCE models with our determined abundance ratios [El/\ion{Fe}{I}] in relation to [\ion{Fe}{I}/H] and age. For this purpose we selected GCE models developed by \citet{Romano2010}, \citet{Kobayashi2011}, and \citet{Prantzos2018}. These studies explicitly provide the GCE in the context of metallicity and age for all or majority of species analysed in this work.
In the case of \citet{Romano2010}, we took their 1st (hereafter R1) and 15th (hereafter R15) models, which represent two completely different model inputs. Model R1 is based on adopted yields from \cite{Woosley1995} case A and \cite{Hoek1997}. Model R15 is computed using the yields from \cite{Kobayashi2006} ($\varepsilon_{\rm{HN}}=1$), the Geneva-group \citep{Ekstrom2008} presupernovae yields computed with rotation and mass loss and the yields from low- and intermediate-mass stars \citep{Karakas2010}. From the \citet{Kobayashi2011} we took the solar neighbourhood thin-disc model (hereafter K$_{\rm{thin}}$ -- magenta line) with the recomputed yields and updated star formation rates (compared to the \citealt{Kobayashi2006} model). Finally, we plotted the baseline model (hereafter PB) from \citet{Prantzos2018}, which for the first time includes, among other factors, the combined effect of metallicity, mass loss, and rotation. We show cases where all models completely fail to represent the data in Figures~\ref{fig:ELEMENTS_separation_failed} and \ref{fig:ELEMENTS_age_failed}, whereas all plots are presented in Figures~\ref{fig:ELEMENTS_separation} and \ref{fig:ELEMENTS_age}. We also discuss the abundances of elements for the \parametersfull~stars in our sample and compare them with the selected models.

\subsection{Mg, Si, Ca, Sc, and Ti}

\begin{table*}
\caption{Best linear fits to the [El/Fe]$-$age distributions in dex\,Gyr$^{-1}$.} 
\begin{tabular}{lccc}
\hline \hline
\smallskip
Elements & This work  & Spina et al. 2016 & Nissen et al. 2017 \\
  & Thin-disc & Solar twins &  Solar twins  \\
\hline
${\rm[\ion{Na}{I}/\ion{Fe}{I}]}$	&	$-0.006	\pm	0.003$	&	$0.0150	\pm	0.0030$	&	$0.0089	\pm	0.0025$	\\
${\rm[\ion{Mg}{I}/\ion{Fe}{I}]}$	&	$0.016	\pm	0.002$	&	$0.0100	\pm	0.0009$	&	$0.0095	\pm	0.0007$	\\
${\rm[\ion{Al}{I}/\ion{Fe}{I}]}$	&	$0.019	\pm	0.003$	&	$0.0147	\pm	0.0015$	&	$0.0148	\pm	0.0011$	\\
${\rm[\ion{Si}{I}/\ion{Fe}{I}]}$	&	$0.009	\pm	0.002$	&	$0.0051	\pm	0.0008$	&	$0.0037	\pm	0.0007$	\\
${\rm[\ion{Si}{II}/\ion{Fe}{I}]}$	&	$0.011	\pm	0.002$		&	&	\\							
${\rm[\ion{S}{I}/\ion{Fe}{I}]}$	&	$0.004	\pm	0.003$	&	$0.0046	\pm	0.0017$	&	$0.0046	\pm	0.0010$	\\
${\rm[\ion{K}{I}/\ion{Fe}{I}]}$	&	$-0.027	\pm	0.006$		&	&	\\							
${\rm[\ion{Ca}{I}/\ion{Fe}{I}]}$	&	$0.001	\pm	0.002$	&	$-0.0014	\pm	0.0008$	&	$-0.0008	\pm	0.0003$	\\
${\rm[\ion{Ca}{II}/\ion{Fe}{I}]}$	&	$0.005	\pm	0.003$	&	&	\\							
${\rm[\ion{Sc}{I}/\ion{Fe}{I}]}$	&	$0.000	\pm	0.002$	&	$0.0071	\pm	0.0018$	&				\\
${\rm[\ion{Sc}{II}/\ion{Fe}{I}]}$	&	$0.004	\pm	0.003$	&	&	\\							
${\rm[\ion{Ti}{I}/\ion{Fe}{I}]}$	&	$0.002	\pm	0.003$	&	$0.0061	\pm	0.0011$	&	$0.0054	\pm	0.0006$	\\
${\rm[\ion{Ti}{II}/\ion{Fe}{I}]}$	&	$0.007	\pm	0.002$	&	&	\\							
${\rm[\ion{V}{I}/\ion{Fe}{I}]}$	&	$-0.005	\pm	0.002$	&	$-0.0034	\pm	0.0012$	&				\\
${\rm[\ion{Cr}{I}/\ion{Fe}{I}]}$	&	$0.001	\pm	0.002$	&	$-0.0024	\pm	0.0005$	&	$-0.0013	\pm	0.0005$	\\
${\rm[\ion{Cr}{II}/\ion{Fe}{I}]}$	&	$-0.003	\pm	0.002$	&	&	\\							
${\rm[\ion{Mn}{I}/\ion{Fe}{I}]}$	&	$-0.010	\pm	0.002$	&	$0.0075	\pm	0.0019$	&				\\
${\rm[\ion{Co}{I}/\ion{Fe}{I}]}$	&	$0.005	\pm	0.003$	&	$0.0120	\pm	0.0030$	&				\\
${\rm[\ion{Ni}{I}/\ion{Fe}{I}]}$	&	$0.004	\pm	0.002$	&	$0.0076	\pm	0.0017$	&	$0.0058	\pm	0.0012$	\\
${\rm[\ion{Cu}{I}/\ion{Fe}{I}]}$	&	$-0.004	\pm	0.002$	&	$0.0200	\pm	0.0030$	&				\\
${\rm[\ion{Zn}{I}/\ion{Fe}{I}]}$	&	$-0.001	\pm	0.003$	&	$0.0122	\pm	0.0014$	&	$0.0112	\pm	0.0009$	\\
\hline
 \label{tab:Age_GRADIENTS}
\end{tabular}
\end{table*}

We observe distinguishable trends for $\alpha$-elements (magnesium, silicon, calcium, scandium, and titanium) of thick- and thin-disc members with higher and lower $\alpha$-element abundances, respectively (Fig.~\ref{fig:ELEMENTS_separation}). The iron content for the thin-disc members ranges from [\ion{Fe}{I}/H]$\approx-0.55$ to 0.35~dex, whereas the thick-disc members exhibit lower metallicity from [\ion{Fe}{I}/H]$\approx-0.8$ to $-0.2$~dex. The separation is most prominent in the case of magnesium. The sequences of element abundances in the two discs are distinguishable for Si, Ca, Sc, and Ti as well. The trends of our results are in agreement with other studies (e.g. \citealt{Adibekyan2012, Bensby2014, Mikolaitis2014}). The two populations overlap more when looking at the calcium abundances which is in agreement with other studies (e.g. \citealt{Bensby2005, Neves2009}).

In Fig.~\ref{fig:ELEMENTS_age} we follow variations of Mg, Si, Ca, Sc, and Ti with age. We observe positive slopes of [El/\ion{Fe}{I}] in relation to age for Mg and Si and almost no slopes for Ca, Sc, and Ti. A number of works that studied gradients of elemental abundances versus age is very small. However, we compared our values with the studies by \citet{Spina2016} and \citet{Nissen2017}, who analysed the [El/Fe] ratios for elements up to Zn as a function of stellar ages for solar twins whose age range is similar to ours. In Table~\ref{tab:Age_GRADIENTS} we provide slopes of linear fits $\Delta {\rm [El/Fe]}/\Delta {\rm Age}$ and compare them with the results by \citet{Spina2016} and \citet{Nissen2017}. The gradients we obtained for Mg, Si, Sc, and Ti are in general agreement with those in \citet{Spina2016} and \citet{Nissen2017}. In the case of calcium, they provide a small negative slope, whereas we find a small positive slope. However, our slope values and those of \citet{Spina2016} and \citet{Nissen2017} for the Ca versus age relation are barely significant (less than 2$\sigma$). 

{
The GCE models for Sc and Ti completely fail to represent the data as it is shown in Fig.~\ref{fig:ELEMENTS_separation_failed} and Fig.~\ref{fig:ELEMENTS_age_failed}.
Scandium is incorrectly modelled as underabundand by three out of the four models (K$_{\rm{thin}}$ was not available for scandium), due to the deficiency of scandium in all sets of massive star yields of GCE studies (see also \citealt{Kobayashi2006,Kobayashi2011,Nomoto2013}). The PB model has the smallest deviation from our data (about $\approx$0.25~dex), whereas the other models predict very negative [Sc/Fe] abundances for thin-disc ($\approx -0.5$~and less).
The titanium model trends [Ti/Fe] are also very negative. According to the work by \citet{Prantzos2018}, a proper analysis of the physical conditions in which $^{\rm{48}}$Cr is produced should be more carefully investigated because the main isotope of Ti ($^{\rm{48}}$Ti) comes from the decay of $^{\rm{48}}$Cr. 
}

{The computations for Mg, Si, and Ca are significantly more accurate; some models are at the level of the data (Figs.~\ref{fig:ELEMENTS_separation} and \ref{fig:ELEMENTS_age}).
Magnesium is predicted to be significantly less abundant by all models except K$_{\rm{thin}}$. \citet{Prantzos2018} admitted that isotopic abundances obtained for the time of the solar system formation are underproduced with the yields they adopted. The same is true for model R1, which uses the \cite{Woosley1995} yields. Model R15 by \citet{Romano2010}, which includes the yields from \cite{Kobayashi2006}, can help reproduce the Mg evolution better, but still not well enough. On the other hand, the \citet{Kobayashi2011} K$_{\rm{thin}}$ model reproduces the abundances quite well in both [Mg/\ion{Fe}{I}] versus \ion{Fe}{I} and [Mg/\ion{Fe}{I}] versus age distributions. However, all the models that we plot can reproduce the [Mg/\ion{Fe}{I}] trends (both in \ion{Fe}{I} and age diagrams) quite well.
Silicon versus \ion{Fe}{I} is quite well reproduced by the PB and R1 models, whereas R15 and K$_{\rm{thin}}$ tend to overestimate the silicon-to-iron ratio; this ratio is reproduced very well versus age, except for the K$_{\rm{thin}}$ model.
Calcium is reproduced better with the PB and K$_{\rm{thin}}$ models in both plots. All four models are predict an increase in [Ca/\ion{Fe}{I}] towards older ages, which agrees with our thin-disc results.}

\subsection{Na and Al}

The sodium abundance seems to have a mild trend that decreases until [\ion{Fe}{I}/H]$\approx-0.2,$ and then goes back up for the solar and metal-rich stars (Fig.~\ref{fig:ELEMENTS_separation}). We did not observe a significant distinction between thin- and thick-disc stars. 
As expected, the trend of  [\ion{Al}{I}/\ion{Fe}{I}] is similar to those found for the $\alpha$-elements.
The  distributions are clearly distinct, with the thick-disc aluminium abundances generally being higher than  the thin-disc values. There is a significant scatter in both thick- and thin-disc sequences. 
In general, the behaviour of Na and Al are similar to those obtained by \citet{Neves2009, Adibekyan2012}, and \citet{Bensby2014}.

For sodium-to-iron abundance ratios we obtain a slightly negative slope in comparison with ages (Fig.~\ref{fig:ELEMENTS_age} and Table~\ref{tab:Age_GRADIENTS}),  whereas \citet{Spina2016} and \citet{Nissen2017} showed a positive slope. For aluminium, in agreement with \citet{Spina2016} and \citet{Nissen2017}, we find clear positive slopes that are similar to those for magnesium.

Sodium is predicted to be underabundant by the R1 and PB models, whereas R15 (which includes the \citealt{Kobayashi2006} yields) agrees better with the data. Interestingly, the K$_{\rm{thin}}$ model is significantly overabundant. The PB model shows a slightly negative abundance trend in relation to age, as can be seen in our data as well.
Aluminium is similarly well predicted by the R1 and R15 models,  whereas the PB model is underabundant by $\approx$0.2~dex and the K$_{\rm{thin}}$ is overabundant in comparison with our data. However, while being overabundant, only K$_{\rm{thin}}$ shows a similar positive trend of  Al abundance versus age  as determined in this work and in \citet{Spina2016}.

\subsection{S and K}

In [\ion{S}{I}/\ion{Fe}{I}] versus [\ion{Fe}{I}/H] space we observe that sulphur abundances behave similarly to $\alpha$-elements. This trend is also seen in \citet{Mishenina2015}. Potassium abundances show larger scatter with some minor decrease towards larger [\ion{Fe}{I}/H]. Generally, the [\ion{K}{I}$_{\rm NLTE}$/\ion{Fe}{I}] versus [\ion{Fe}{I}]/H] behaviour is similar to those reported by \citet{Takeda2002} and \citet{Zhao2016}.

The slope of sulphur distribution versus age is mildly positive and agrees well with the values provided by both \citet{Spina2016} and \citet{Nissen2017}. The [\ion{K}{I}$_{\rm NLTE}$/\ion{Fe}{I}] distribution versus age is slightly positive, but of low significance. \citet{Spina2016} did not provide any slopes for potassium, which seems to have no definite dependence on age, probably because its abundance determinations are based on the large  saturated line at 7698.97~$\AA$ (e.g.
$\approx$154~$m\AA$ for the Sun, \citealt{Moore1966}),  which is more suited for the analysis by  spectral synthesis.

{Potassium abundances (Figs.~\ref{fig:ELEMENTS_separation_failed}~and~\ref{fig:ELEMENTS_age_failed}) are underestimated by all four models (PB, R1, R15, and  K$_{\rm{thin}}$);  the differences are $\approx$0.25~dex. The two models  that are based on \citet{Kobayashi2006} and \citet{Kobayashi2011} (i.e. K$_{\rm{thin}}$,
R15) yields underestimated potassium abundances (by $>0.65$~dex), and they are even out of range in our plots. The PB and R1 models predict a mild positive slope of [\ion{K}{I}$_{\rm NLTE}$/\ion{Fe}{I}] versus  age, which  is also seen in our data.}

{Sulphur behaviour in the thin-disc is better predicted by the PB and K$_{\rm{thin}}$ models (see Figs.~\ref{fig:ELEMENTS_separation}~and~\ref{fig:ELEMENTS_age}). The K$_{\rm{thin}}$ model  most closely reproduces the results. The mild positive slope of abundance ratios versus age is seen by all four models; however, R1 and R15 are underabundant by $\approx$0.2~dex.
}

  \begin{figure}[]%[htb]
  \centering
   \advance\leftskip 0cm
   \includegraphics[scale=0.51]{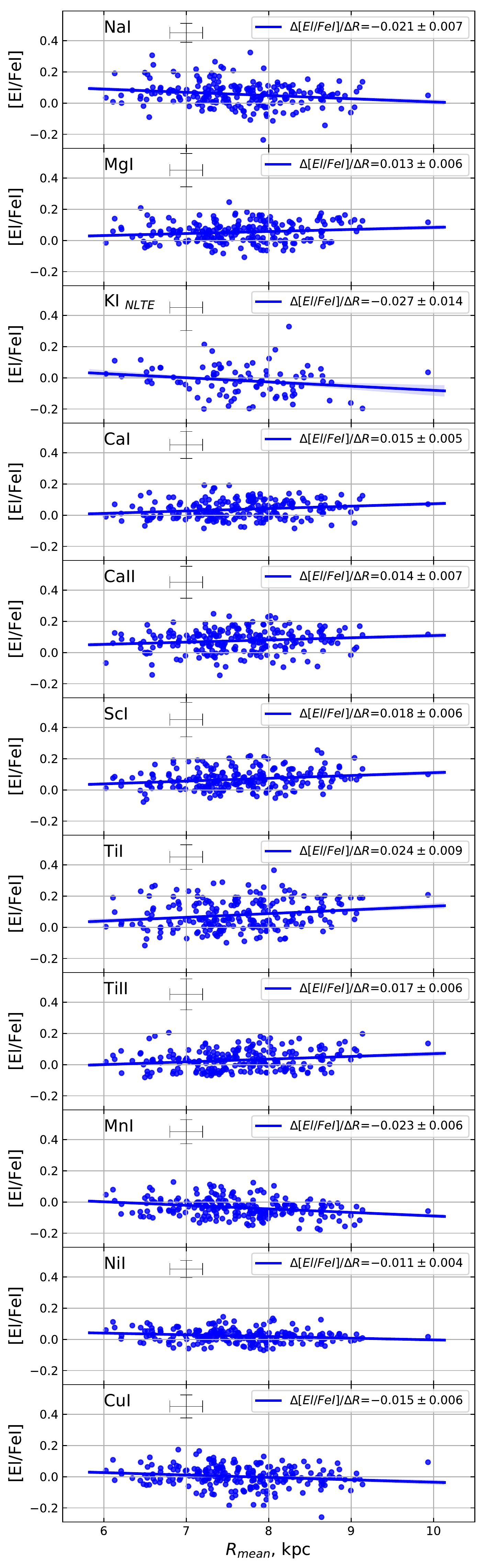}
  \caption{
Observed element-to-iron abundance ratios as a function of $R_{\rm{mean}}$ (significance $> 2~\sigma$). 
  }
  \label{fig:ELEMENTS_rmean}
  \end{figure}

\begin{figure}[]%[htb]
  \centering
   \advance\leftskip 0cm
   \includegraphics[scale=0.51]{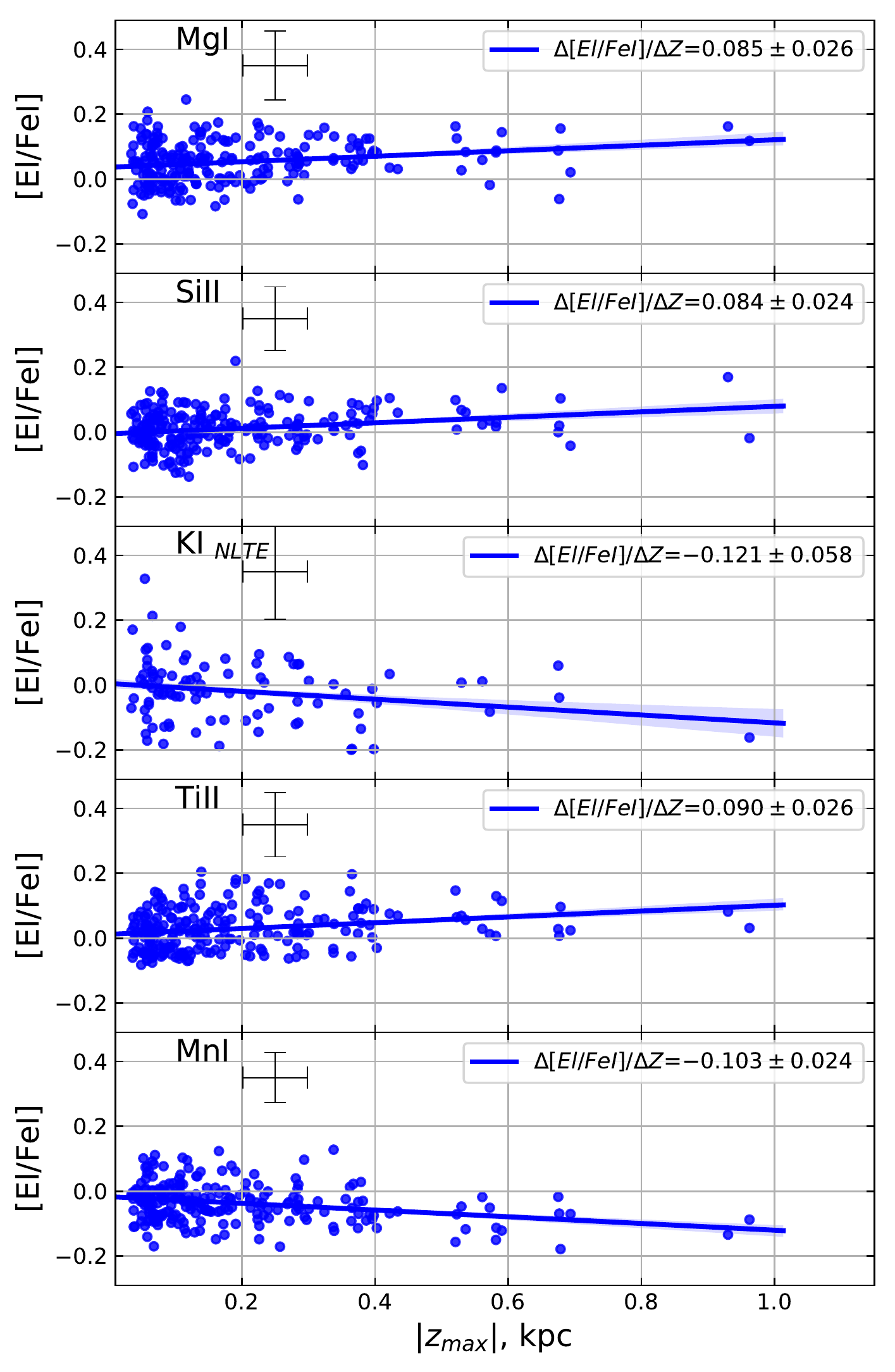}
  \caption{
Observed element-to-iron abundance ratios as a function of $z_{\rm{max}}$ (significance $> 2~\sigma$). 
  }
  \label{fig:ELEMENTS_zmax}
  \end{figure}

\subsection{V, Cr, Co, Ni, Mn}

The V- and Co-to-Fe ratios in the thin- and thick-disc stars resemble those of $\alpha$-elements, whereas Cr and Ni abundances are well confined around the \ion{Fe}{I} value. The manganese-to-iron ratio trends are opposite to those of $\alpha$-elements for  the reasons mentioned above. In general, the LTE abundance trends of these iron-peak elements are similar to those obtained by \citet{Neves2009, Adibekyan2012, Bensby2014}, and \citet{Battistini2015}.

The slopes of [El/\ion{Fe}{I}] versus the age distributions of Cr, V, Co, and Ni are not very steep and in general agreement with those from studies by \citet{Spina2016} and \citet{Nissen2017}. The slope of manganese abundance versus age determined in this work is actually opposite to that of solar twins by \citet{Spina2016}, which is expected as a  majority of the older stars are found at the metal-poor end of the thin-disc where the lowest manganese-to-iron ratios are.

The elements V and Ni are other examples where all models fail to represent the data (Figs.~\ref{fig:ELEMENTS_separation_failed}~and~\ref{fig:ELEMENTS_age_failed}).
Vanadium models underestimate the observed abundances by at least $\approx$0.2~dex. The slight negative slope is predicted by the PB and K$_{\rm{thin}}$ models, which is in agreement with our study and the results from \citet{Spina2016}.
Nickel is overestimated by all the models we use, and the disagreement between the results and theoretical calculations is very large, reaching more than 0.4~dex for the \citet{Romano2010} models and $\approx$0.27~dex for the \citet{Prantzos2018} PB model (with K$_{\rm{thin}}$ being in between). All models predict a slightly negative slope in the [\ion{Ni}{I}/\ion{Fe}{I}] versus age relation, which disagree with  our results and with those by other authors (\citealt{Spina2016} and \citealt{Nissen2017}).

{Chromium is modelled quite well by the PB, R15, and K$_{\rm{thin}}$ models; however, R1 shows a minor overabundance. 
The manganese abundance in the thin-disc is better reproduced by the PB, R15, and K$_{\rm{thin}}$ models. All models predict the negative slope of [\ion{Mn}{I}/\ion{Fe}{I}] versus  age relation.
The results of cobalt favour the R1 model, whereas the other three seem to underestimate its abundance. The slope of [\ion{Mn}{I}/\ion{Fe}{I}] versus age is better recreated by the R1 model. The slope from the PB seems to be negative, while our data and that by \citet{Spina2016} show some positive tendency.}

\subsection{Cu, Zn}

In Fig.~\ref{fig:ELEMENTS_separation} we see that the Galactic components have distinct behaviours in the [\ion{Cu}{I}/\ion{Fe}{I}] versus [\ion{Fe}{I}/H] plane. The thick-disc stars look slightly more enriched in Cu than the thin-disc stars
 and the thin-disc exhibits a shallow increase in [\ion{Cu}{I}/\ion{Fe}{I}] at super-solar metallicities. Our copper trends are similar to those from other studies (\citealt{Bensby2014,Mikolaitis2017,Delgado2017}), 
whereas zinc has an $\alpha$-like behaviour that was already reported in some previous studies (\citealt{Bisterzo2006,Saito2009,Mishenina2011,Bensby2014,Mikolaitis2017,Delgado2017}). As is seen in Fig.~\ref{fig:ELEMENTS_age} and Table~\ref{tab:Age_GRADIENTS}, the copper and zinc distributions display almost no trends with the age, which disagree with the studies of solar twins by \citet{Spina2016} and \citet{Nissen2017}, who report clear positive slopes.

{Copper is better reproduced by the R1 model, whereas R15 and PB underestimate [\ion{Cu}{I}/\ion{Fe}{I}] ratios and K$_{\rm{thin}}$ does the opposite and predicts values that are a bit higher  (Figs.~\ref{fig:ELEMENTS_separation}~and~\ref{fig:ELEMENTS_age}). All models predict negative slopes in relation to age and that is in broad agreement with our data; however, solar twins behave differently \citep{Spina2016}.
}
{Zinc favours the R15 model in  element abundance versus metallicity and versus age relations. R1 predicts the unrealistic underabundance towards lower metallicities and the PB model generally shows a strong underabundance by $\approx$0.25~dex, while K$_{\rm{thin}}$ is slightly overabundant at higher metallicities.}

\section{Radial and vertical abundance gradients in the thin-disc}
\label{sec:gradients}

Figures~\ref{fig:ELEMENTS_rmean}~and~\ref{fig:ELEMENTS_zmax} display the [El/\ion{Fe}{I}] significant gradients ($> 2~\sigma$)  in relation to the mean Galactocentric distance ($R_{\rm mean}$) and the maximum height above the Galactic plane ($|z_{\rm max}|$) in the thin-disc sample, while all numerical values of the gradients are presented in Table~\ref{tab:gradients}.
% The values of the slopes $\Delta [El/Fe]/\Delta R_{\rm mean}$ are provided inside the plots. 
We find positive slopes ($> 2~\sigma$) for most of the $\alpha$-elements: \ion{Mg}{I}, \ion{Ca}{I}, \ion{Ca}{II},\ion{Sc}{I}, \ion{Ti}{I}, \ion{Ti}{II}. 
% However, \ion{Si}{I}, \ion{Si}{II}, \ion{Sc}{II} do not show significant abundance trends. 
On the other hand, \ion{Na}{I}, \ion{Mn}{I}, \ion{Ni}{I}, and \ion{Cu}{I} show a significant decrease in abundance with increasing Galactocentric distances.

\begin{table}
\caption{Gradients in relation to the mean Galactocentric distance ($R_{\rm mean}$) and the maximum height above the Galactic plane ($|z_{\rm max}|$) in the thin-disc sample.}
%\vspace{0.2cm}
\centering
\begin{tabular}{lrrr}
%\smallskip
\hline\hline
Element &  $\Delta [X/Fe]/\Delta R_{\rm mean}$ & $\Delta [X/Fe]/\Delta z_{\rm max}$ \\
                              & dex/kpc                        & dex/kpc   \\
\hline

\ion{Na}{I}  	&	 $ -0.021 \pm 0.007 $ 	&	 $ -0.024 \pm 0.032 $ \\
\ion{Mg}{I}  	&	 $ 0.013 \pm 0.006 $ 	&	 $ 0.085 \pm 0.026 $ \\
\ion{Al}{I}  	&	 $ -0.007 \pm 0.008 $ 	&	 $ 0.061 \pm 0.034 $ \\
\ion{Si}{I}  	&	 $ 0.002 \pm 0.005 $ 	&	 $ 0.040 \pm 0.021 $ \\
\ion{Si}{II}  	&	 $ 0.009 \pm 0.006 $ 	&	 $ 0.084 \pm 0.024 $ \\
\ion{S}{I}  	&	 $ -0.003 \pm 0.008 $ 	&	 $ 0.001 \pm 0.035 $ \\
\ion{K}{I}$_{NLTE}$& $ -0.027 \pm 0.014 $ 	&	 $ -0.121 \pm 0.058 $ \\
\ion{Ca}{I}  	&	 $ 0.015 \pm 0.005 $ 	&	 $ -0.002 \pm 0.024 $ \\
\ion{Ca}{II}  	&	 $ 0.014 \pm 0.007 $ 	&	 $ 0.039 \pm 0.030 $ \\
\ion{Sc}{I}  	&	 $ 0.018 \pm 0.006 $ 	&	 $ 0.012 \pm 0.026 $ \\
\ion{Sc}{II}  	&	 $ 0.006 \pm 0.007 $ 	&	 $ 0.110 \pm 0.028 $ \\
\ion{Ti}{I}  	&	 $ 0.024 \pm 0.009 $ 	&	 $ 0.014 \pm 0.038 $ \\
\ion{Ti}{II}  	&	 $ 0.017 \pm 0.006 $ 	&	 $ 0.090 \pm 0.026 $ \\
\ion{V}{I}  	&	 $ 0.002 \pm 0.005 $ 	&	 $ -0.017 \pm 0.022 $ \\
\ion{Cr}{I}  	&	 $ 0.007 \pm 0.004 $ 	&	 $ -0.008 \pm 0.018 $ \\
\ion{Cr}{II}  	&	 $ 0.008 \pm 0.006 $ 	&	 $ 0.009 \pm 0.026 $ \\
\ion{Mn}{I}  	&	 $ -0.023 \pm 0.006 $ 	&	 $ -0.103 \pm 0.024 $ \\
\ion{Co}{I}  	&	 $ 0.000 \pm 0.007 $ 	&	 $ 0.060 \pm 0.029 $ \\
\ion{Ni}{I}  	&	 $ -0.011 \pm 0.004 $ 	&	 $ 0.013 \pm 0.017 $ \\
\ion{Cu}{I}  	&	 $ -0.015 \pm 0.006 $ 	&	 $ 0.030 \pm 0.028 $ \\
\ion{Zn}{I}  	&	 $ 0.004 \pm 0.007 $ 	&	 $ 0.039 \pm 0.031 $ \\

\hline
\\
\end{tabular}
%\begin{tablenotes}
\\
%\end{tablenotes}
 \label{tab:gradients}
\end{table} 

When looking at the $\Delta [X/Fe]/\Delta z_{\rm max}$ relations, the situation is similar. The $\alpha$-elements generally have positive slopes with \ion{Mg}{I}, \ion{Si}{II}, \ion{Sc}{II}, and \ion{Ti}{II} being significant by more than $> 2~\sigma$,  whereas only \ion{Mn}{I} shows a significant negative trend ($4.2~\sigma$).  
The  case of \ion{K}{I}$_{\rm NLTE}$ is interesting: it shows negative slopes  in relation to $R_{\rm mean}$ and to $|z_{max}|$, which are significant by $\approx2~\sigma$. 

{There are no published data on [El/Fe] gradients in relation to $R_{\rm mean}$ or $|z_{max}|$. Typically, elemental abundance gradients are studied by large spectroscopic surveys that observe more distant stars, and they usually study gradients according to the current $R_{GC}$ and $|z|$ or the the guiding radii ($R_{\rm g}$). However, even large spectroscopic surveys have published very few elemental radial or vertical gradients for field stars separated into thin- or thick-disc components. }

\begin{table*}
\caption{Radial and vertical [El/Fe] gradients in the Galactic thin-disc from literature.} 
\begin{tabular}{lcccccc}
\hline \hline
\smallskip
Elements & Value (dex~kpc$^{-1}$)  & Source & Deffinition of thin-disc & Comments \\
\hline \hline
Galactocentric\\
\hline
$\Delta [\alpha/{\rm Fe}]/\Delta R_{\rm g}$	&	$-0.004	\pm	0.001$	&	\citet{Boeche2013} & $|Z_{\rm max}| < 0.04$~kpc	&	RAVE sample	\\
$\Delta [\alpha/{\rm Fe}]/\Delta R_{\rm g}$	&	$ 0.010	\pm	0.002$	&	\citet{Boeche2013} & $|Z_{\rm max}| < 0.04$~kpc	&	GCS sample	\\
$\Delta [\alpha/{\rm Fe}]/\Delta R_{\rm GC}$	&	$-0.005	\pm	0.002$	&	\citet{Anders2014} & $0.0<|z|<0.4$~kpc      	&	APOGEE gold sample	\\
$\Delta$[\ion{Mg}{I}/{\rm Fe}]/$\Delta R_{\rm GC}$&	$0.021	\pm	0.016$&	\citet{Bergemann2014} & $0.0<|z|<0.3$~kpc      	&	GES UVES	\\
$\Delta$[\ion{Mg}{I}/{\rm Fe}]/$\Delta R_{\rm GC}$&	$0.009	\pm	0.003$&	\citet{Mikolaitis2014} & Chemical      	&	GES GIRAFFE	\\
$\Delta$[\ion{Si}{I}/{\rm Fe}]/$\Delta R_{\rm GC}$&	$0.021	\pm	0.004$&	\citet{Mikolaitis2014} & Chemical      	&	GES GIRAFFE	\\
$\Delta$[\ion{Ca}{II}/{\rm Fe}]/$\Delta R_{\rm GC}$&	$0.017	\pm	0.004$&	\citet{Mikolaitis2014} & Chemical      	&	GES GIRAFFE	\\
$\Delta$[Na/Fe]/$\Delta R_{\rm GC}$ & 0.007$\pm$0.002 & \citet{Genovali2015} &  & Cepheids\\ 
$\Delta$[Na/Fe]/$\Delta R_{\rm GC}$ & 0.001$\pm$0.002 & \citet{Genovali2015} &  & Cepheids\\
$\Delta$[Mg/Fe]/$\Delta R_{\rm GC}$ & 0.013$\pm$0.003 & \citet{Genovali2015} &  & Cepheids\\
$\Delta$[Si/Fe]/$\Delta R_{\rm GC}$ & 0.007$\pm$0.002 & \citet{Genovali2015} &  & Cepheids\\
$\Delta$[Ca/Fe]/$\Delta R_{\rm GC}$ & 0.018$\pm$0.002 & \citet{Genovali2015} &  & Cepheids\\
\hline
Vertical\\
\hline
$\Delta$[\ion{Mg}{I}/Fe]$\Delta z$ & 0.041$\pm$0.004 & \citet{Mikolaitis2014} & Chemical & GES GIRAFFE\\ 
$\Delta$[\ion{Al}{I}/Fe]$\Delta z$ & 0.051$\pm$0.004 & \citet{Mikolaitis2014} & Chemical & GES GIRAFFE\\ 
$\Delta$[\ion{Si}{I}/Fe]$\Delta z$ & 0.006$\pm$0.008 & \citet{Mikolaitis2014} & Chemical & GES GIRAFFE\\ 
$\Delta$[\ion{Ca}{II}/Fe]$\Delta z$ &  0.034$\pm$0.010 & \citet{Mikolaitis2014} & Chemical & GES GIRAFFE\\ 
$\Delta$[$\alpha$/Fe]/$\Delta z$ & 0.008$\pm$0.002 & \citet{Duong2018} & Chemical and Kinematical & GALAH survey\\ 
$\Delta$[$\alpha$/Fe]/$\Delta z$ & 0.022$\pm$0.001 & \citet{Li2018} & Kinematical & GES GIRAFFE\\

\hline
 \label{tab:gradients_literature}
\end{tabular}
\end{table*}

% \citet{Boeche2013} derived radial slopes for magnesium, silicon, and aluminium
%  from the RAVE and the Geneva-Copenhagen (GCS) survey samples as a function of $|Z_{\rm max}|$.  
% They found negative $\Delta [\alpha/{\rm Fe}]/\Delta R_{\rm g}=-0.004\pm0.001$~dex\,kpc$^{-1}$ slopes for their RAVE sample in the $|Z_{\rm max}| < 0.04$~kpc range where the thin-disc dominates. Interestingly, they found a positive $\Delta [\alpha/{\rm Fe}]/\Delta R_{\rm g}=0.010\pm0.002$~dex\,kpc$^{-1}$ for the GCS sample in the same $|z_{\rm max}| < 0.04$~kpc range.
% From the APOGEE survey data, \citet{Anders2014} found a flat $\Delta [\alpha/{\rm Fe}]/\Delta R_\rm{GC}$ gradient in the thin-disc-dominated range $0.0<|z|<0.4$~kpc.
% The UVES-based GES study by \citet{Bergemann2014} measured the magnesium gradient $\Delta$[\ion{Mg}{I}/{\rm Fe}]/$\Delta R_{\rm GC}= 0.021\pm0.016$~dex\,kpc$^{-1}$ for $|z|<0.3$~kpc.
% For the GES GIRAFFE-based work by \citet{Mikolaitis2014} generally positive gradients were determined for the thin-disc sample: $\Delta$[\ion{Mg}{I}/{\rm Fe}]/$\Delta R_{\rm GC}= 0.009\pm0.003$~dex~kpc$^{-1}$, $\Delta$[\ion{Si}{I}/{\rm Fe}]/$\Delta R_{\rm GC}= 0.021\pm0.004$~dex~kpc$^{-1}$, $\Delta$[\ion{Ca}{II}/{\rm Fe}]/$\Delta R_{\rm GC}= 0.017\pm0.003$~dex~kpc$^{-1}$. 
% %
The data of RAVE, APOGEE, and GES are quite different from our (different target selection, different distance estimation methods etc.); however, the gradients determined by  \citet{Boeche2013}, \citet{Bergemann2014}, and \citet{Mikolaitis2014} qualitatively agree with ours (Table~\ref{tab:gradients_literature}). The thin-disc abundance gradients can be compared with the studies based on Cepheid variables as well.
For example, slightly positive gradients of individual $\alpha$-elements seem to agree with the work by \citet{Genovali2015}. 
Since the Cepheids are young metal-rich stars, they show the present-day gradients (\citealt{Cescutti2007}), thus we can only search for a broad agreement to our results.

{Unfortunately, there are few elemental abundance vertical gradients reported in the literature, and they are mostly based on $\alpha$-elements.
The GES GIRAFFE-based study by \citet{Mikolaitis2014} and \citet{Li2018} from the APOGEE survey obtained generally positive gradients in the thin-disc sample, while the GALAH survey (\citealt{Duong2018}) determined only a shallow positive gradient of $\Delta$[$\alpha$/Fe]/$\Delta z$.}

\section{Summary}
\label{sec:summary}

This paper is the second release of spectroscopic data from the Spectroscopic and Photometric Survey of the Northern Sky (SPFOT). The survey aims to provide a detailed chemical composition from high-resolution spectra and photometric variability data for bright stars in the northern sky using telescopes at the Molėtai Astronomical Observatory, Vilnius University.

We observed high-resolution spectra for all \allnpf~photometrically selected 6500~K and cooler dwarfs in the sky area with radius of 20~degrees centred on the preliminary ESA-PLATO NPF field and derived the main atmospheric parameters for \parametersnpf~stars. We co-added this sample with  \parametersstep~stars from the STEP~2 field investigated in \citetalias{Mikolaitis2018} and computed kinematic parameters, ages, and abundances of 21~chemical species for \parametersfull~FGK~stars in total.
These regions of the sky are becoming very important as they will soon be intensively observed by the NASA TESS mission (\citealt{Ricker2015,Sullivan2015}).

We turned our attention to bright stars since only up to 30\% of  the stars that are brighter than  8~mag  have some previous spectroscopic observations. Apart from faint and distant stars, which are more often observed by large spectroscopic surveys, it is important to pay attention to the bright nearby stars as well, especially for the objects in the ``hot'' regions like the TESS continuous viewing zones (\citealt{Sharma2018}), and possible PLATO fields (\citealt{Miglio2017}).

With this study we also examined whether bright solar vicinity stars can provide important information about the Galactic chemical evolution. 
We used the [Mg/Fe] ratio to divide the sample stars to the thin-disc (223 stars) and thick-disc (26 stars). We then explored the behaviour of 21 chemical species in [El/\ion{Fe}{I}] versus [\ion{Fe}{I}/H] and [El/\ion{Fe}{I}] versus age planes. We restricted our conclusions to the thin-disc stars since the sample of thick-disc stars still should be increased. Thus, for the thin-disc stars we found that $\alpha$-elements and aluminium abundances have positive trends with age, while the trend of the Mn abundance is obviously negative.

In order to better understand the chemical evolution of the solar neighbourhood, we compared our observational results to the recent theoretical models 
by \citet{Romano2010, Kobayashi2011}, and \citet{Prantzos2018} that provide predictions for a variety of chemical elements (up to $Z=30$) not only in the [El/\ion{Fe}{I}] versus [\ion{Fe}{I}/H] plane, but also in the [El/\ion{Fe}{I}] versus age plane.
The comparison between the observed data and the theoretical models led to the conclusion that it is still difficult for scientists to model the nucleosynthesis channels for potasium, scandium, titanium, vanadium, and nickel. 
Even if adjustments in some metallicity domains are still required, other elements are modelled rather well. 

In our study, we attempted to estimate elemental abundance ([El/\ion{Fe}{I}]) gradients versus the mean Galactocentric distance and height above the Galactic plane. Our stars are in close proximity to the Sun ($7.9<R_{\rm GC}<8.2$~kpc, $0.02<|z|<0.14$~kpc); however, their orbital parameters revealed quite wide ranges of their mean Galactocentric distances and maximum heights above the Galactic plane ($5.91<R_{\rm mean}<10.51$~kpc, $0.03<|z_{\rm max}|<1.29$~kpc). We prove that using $R_{\rm mean}$ and $|z_{\rm max}|$ is very helpful while estimating the gradients of element abundance ratios in the Galactic disc. 
We found that the $\alpha$-element and zinc abundances have slightly positive or flat radial and vertical gradients, while gradients for the odd-{\it Z} element Na, K, V, and Mn abundances are negative.

The study of vertical and radial abundance gradients in the Galactic discs, as well as the age dependence of element abundances can provide strong evidence of the mechanisms of Galaxy formation and should be continued.

\begin{acknowledgements}

This research has made use of the SIMBAD database and
NASA Astrophysics Data System (operated at CDS, Strasbourg, France).
This work has made use of data from the European Space Agency (ESA) mission
{\it Gaia} (\url{https://www.cosmos.esa.int/gaia}), processed by the {\it Gaia}
Data Processing and Analysis Consortium (DPAC,
\url{https://www.cosmos.esa.int/web/gaia/dpac/consortium}). Funding for the DPAC
has been provided by national institutions, in particular the institutions
participating in the {\it Gaia} Multilateral Agreement. We are especially grateful to 
T. Masseron and B. Plez for providing us with  molecular data.
We appreciate that D.~Romano, N. Prantzos, C.~I.~Johnson, and C.~Kobayashi kindly shared their model data.
This research was funded by the grant from the Research Council of Lithuania (LAT-08/2016).
\end{acknowledgements}

\bibliographystyle{aa} % style aa.bst

\bibliography{SPOT-II.bbl}        % PLATO.bib is the name of our database

\newcommand{\noop}[1]{}
\begin{thebibliography}{107}
\expandafter\ifx\csname natexlab\endcsname\relax\def\natexlab#1{#1}\fi

\bibitem[{{Adibekyan} {et~al.}(2012){Adibekyan}, {Sousa}, {Santos}, {Delgado
  Mena}, {Gonz{\'a}lez Hern{\'a}ndez}, {Israelian}, {Mayor}, \&
  {Khachatryan}}]{Adibekyan2012}
{Adibekyan}, V.~Z., {Sousa}, S.~G., {Santos}, N.~C., {et~al.} 2012, \aap, 545,
  A32

\bibitem[{{Alexeeva} {et~al.}(2018){Alexeeva}, {Ryabchikova}, {Mashonkina}, \&
  {Hu}}]{Alexeeva2018}
{Alexeeva}, S., {Ryabchikova}, T., {Mashonkina}, L., \& {Hu}, S. 2018, \apj,
  866, 153

\bibitem[{{Alvarez} \& {Plez}(1998)}]{Alvarez1998}
{Alvarez}, R. \& {Plez}, B. 1998, \aap, 330, 1109

\bibitem[{{Anders} {et~al.}(2014){Anders}, {Chiappini}, {Santiago},
  {Rocha-Pinto}, {Girardi}, {da Costa}, {Maia}, {Steinmetz}, {Minchev},
  {Schultheis}, {Boeche}, {Miglio}, {Montalb{\'a}n}, {Schneider}, {Beers},
  {Cunha}, {Allende Prieto}, {Balbinot}, {Bizyaev}, {Brauer}, {Brinkmann},
  {Frinchaboy}, {Garc{\'{\i}}a P{\'e}rez}, {Hayden}, {Hearty}, {Holtzman},
  {Johnson}, {Kinemuchi}, {Majewski}, {Malanushenko}, {Malanushenko},
  {Nidever}, {O'Connell}, {Pan}, {Robin}, {Schiavon}, {Shetrone}, {Skrutskie},
  {Smith}, {Stassun}, \& {Zasowski}}]{Anders2014}
{Anders}, F., {Chiappini}, C., {Santiago}, B.~X., {et~al.} 2014, \aap, 564,
  A115

\bibitem[{{Andrievsky} {et~al.}(2008){Andrievsky}, {Spite}, {Korotin}, {Spite},
  {Bonifacio}, {Cayrel}, {Hill}, \& {Fran{\c c}ois}}]{Andrievsky2008}
{Andrievsky}, S.~M., {Spite}, M., {Korotin}, S.~A., {et~al.} 2008, \aap, 481,
  481

\bibitem[{{Battistini} \& {Bensby}(2015)}]{Battistini2015}
{Battistini}, C. \& {Bensby}, T. 2015, \aap, 577, A9

\bibitem[{{Baumueller} \& {Gehren}(1997)}]{Baumueller1997}
{Baumueller}, D. \& {Gehren}, T. 1997, \aap, 325, 1088

\bibitem[{{Bensby} {et~al.}(2003){Bensby}, {Feltzing}, \&
  {Lundstr{\"o}m}}]{Bensby2003}
{Bensby}, T., {Feltzing}, S., \& {Lundstr{\"o}m}, I. 2003, \aap, 410, 527

\bibitem[{{Bensby} {et~al.}(2005){Bensby}, {Feltzing}, {Lundstr{\"o}m}, \&
  {Ilyin}}]{Bensby2005}
{Bensby}, T., {Feltzing}, S., {Lundstr{\"o}m}, I., \& {Ilyin}, I. 2005, \aap,
  433, 185

\bibitem[{{Bensby} {et~al.}(2014){Bensby}, {Feltzing}, \& {Oey}}]{Bensby2014}
{Bensby}, T., {Feltzing}, S., \& {Oey}, M.~S. 2014, \aap, 562, A71

\bibitem[{{Bergemann}(2011)}]{Bergemann2011}
{Bergemann}, M. 2011, \mnras, 413, 2184

\bibitem[{{Bergemann} \& {Gehren}(2008)}]{Bergemann2008}
{Bergemann}, M. \& {Gehren}, T. 2008, \aap, 492, 823

\bibitem[{{Bergemann} {et~al.}(2013){Bergemann}, {Kudritzki}, {W{\"u}rl},
  {Plez}, {Davies}, \& {Gazak}}]{Bergemann2013}
{Bergemann}, M., {Kudritzki}, R.-P., {W{\"u}rl}, M., {et~al.} 2013, \apj, 764,
  115

\bibitem[{{Bergemann} {et~al.}(2014){Bergemann}, {Ruchti}, {Serenelli},
  {Feltzing}, {Alves-Brito}, {Asplund}, {Bensby}, {Gruyters}, {Heiter},
  {Hourihane}, {Korn}, {Lind}, {Marino}, {Jofre}, {Nordlander}, {Ryde},
  {Worley}, {Gilmore}, {Randich}, {Ferguson}, {Jeffries}, {Micela},
  {Negueruela}, {Prusti}, {Rix}, {Vallenari}, {Alfaro}, {Allende Prieto},
  {Bragaglia}, {Koposov}, {Lanzafame}, {Pancino}, {Recio-Blanco}, {Smiljanic},
  {Walton}, {Costado}, {Franciosini}, {Hill}, {Lardo}, {de Laverny}, {Magrini},
  {Maiorca}, {Masseron}, {Morbidelli}, {Sacco}, {Kordopatis}, \& {Tautvai{\v
  s}ien{\.e}}}]{Bergemann2014}
{Bergemann}, M., {Ruchti}, G.~R., {Serenelli}, A., {et~al.} 2014, \aap, 565,
  A89

\bibitem[{{Bergstr{\"o}m} {et~al.}(1989){Bergstr{\"o}m}, {Peng}, \&
  {Persson}}]{Bergstrom1989}
{Bergstr{\"o}m}, H., {Peng}, W.~X., \& {Persson}, A. 1989, Zeitschrift fur
  Physik D Atoms Molecules Clusters, 13, 203

\bibitem[{{Bisterzo} {et~al.}(2006){Bisterzo}, {Gallino}, \&
  {Pignatari}}]{Bisterzo2006}
{Bisterzo}, S., {Gallino}, R., \& {Pignatari}, M. 2006, {Cu and Zn in
  Thick-Disk and Thin-Disk Stars}, ed. S.~{Randich} \& L.~{Pasquini}, 39

\bibitem[{{Blackwell-Whitehead} {et~al.}(2005){Blackwell-Whitehead},
  {Pickering}, {Pearse}, \& {Nave}}]{Blackwell-Whitehead2005a}
{Blackwell-Whitehead}, R.~J., {Pickering}, J.~C., {Pearse}, O., \& {Nave}, G.
  2005, \apjs, 157, 402

\bibitem[{{Boeche} {et~al.}(2013){Boeche}, {Siebert}, {Piffl}, {Just},
  {Steinmetz}, {Sharma}, {Kordopatis}, {Gilmore}, {Chiappini}, {Williams},
  {Grebel}, {Bland-Hawthorn}, {Gibson}, {Munari}, {Siviero}, {Bienaym{\'e}},
  {Navarro}, {Parker}, {Reid}, {Seabroke}, {Watson}, {Wyse}, \&
  {Zwitter}}]{Boeche2013}
{Boeche}, C., {Siebert}, A., {Piffl}, T., {et~al.} 2013, \aap, 559, A59

\bibitem[{{Bovy}(2015)}]{Bovy15}
{Bovy}, J. 2015, \apjs, 216, 29

\bibitem[{{Bovy} {et~al.}(2012){Bovy}, {Allende Prieto}, {Beers}, {Bizyaev},
  {da Costa}, {Cunha}, {Ebelke}, {Eisenstein}, {Frinchaboy}, {Garc{\'{\i}}a
  P{\'e}rez}, {Girardi}, {Hearty}, {Hogg}, {Holtzman}, {Maia}, {Majewski},
  {Malanushenko}, {Malanushenko}, {M{\'e}sz{\'a}ros}, {Nidever}, {O'Connell},
  {O'Donnell}, {Oravetz}, {Pan}, {Rocha-Pinto}, {Schiavon}, {Schneider},
  {Schultheis}, {Skrutskie}, {Smith}, {Weinberg}, {Wilson}, \&
  {Zasowski}}]{Bovy2012}
{Bovy}, J., {Allende Prieto}, C., {Beers}, T.~C., {et~al.} 2012, \apj, 759, 131

\bibitem[{Bressan {et~al.}(2012)Bressan, Marigo, Girardi, Salasnich, Dal~Cero,
  Rubele, \& Nanni}]{Bressan12}
Bressan, A., Marigo, P., Girardi, L., {et~al.} 2012, Monthly Notices of the
  Royal Astronomical Society, 427, 127

\bibitem[{{Brodzinski} {et~al.}(1987){Brodzinski}, {Kronfeldt}, {Kropp}, \&
  {Winkler}}]{Brodzinski1987}
{Brodzinski}, T., {Kronfeldt}, H.-D., {Kropp}, J.-R., \& {Winkler}, R. 1987,
  Zeitschrift fur Physik D Atoms Molecules Clusters, 7, 161

\bibitem[{{Brooke} {et~al.}(2013){Brooke}, {Bernath}, {Schmidt}, \&
  {Bacskay}}]{Brooke2013}
{Brooke}, J.~S.~A., {Bernath}, P.~F., {Schmidt}, T.~W., \& {Bacskay}, G.~B.
  2013, \jqsrt, 124, 11

\bibitem[{{Casagrande} {et~al.}(2011){Casagrande}, {Sch{\"o}nrich}, {Asplund},
  {Cassisi}, {Ram{\'{\i}}rez}, {Mel{\'e}ndez}, {Bensby}, \&
  {Feltzing}}]{Casagrande2011}
{Casagrande}, L., {Sch{\"o}nrich}, R., {Asplund}, M., {et~al.} 2011, \aap, 530,
  A138

\bibitem[{{Cescutti} {et~al.}(2007){Cescutti}, {Matteucci}, {Fran{\c c}ois}, \&
  {Chiappini}}]{Cescutti2007}
{Cescutti}, G., {Matteucci}, F., {Fran{\c c}ois}, P., \& {Chiappini}, C. 2007,
  \aap, 462, 943

\bibitem[{{Chen} {et~al.}(2004){Chen}, {Nissen}, \& {Zhao}}]{Chen2004}
{Chen}, Y.~Q., {Nissen}, P.~E., \& {Zhao}, G. 2004, \aap, 425, 697

\bibitem[{{Davis} {et~al.}(1971){Davis}, {Wright}, \& {Balling}}]{Davis1971}
{Davis}, S.~J., {Wright}, J.~J., \& {Balling}, L.~C. 1971, \pra, 3, 1220

\bibitem[{{Delgado Mena} {et~al.}(2017){Delgado Mena}, {Tsantaki}, {Adibekyan},
  {Sousa}, {Santos}, {Gonz{\'a}lez Hern{\'a}ndez}, \&
  {Israelian}}]{Delgado2017}
{Delgado Mena}, E., {Tsantaki}, M., {Adibekyan}, V.~Z., {et~al.} 2017, \aap,
  606, A94

\bibitem[{{Dembczy{\'n}ski} {et~al.}(1979){Dembczy{\'n}ski}, {Ertmer},
  {Johann}, {Penselin}, \& {Stinner}}]{Dembczynski1979}
{Dembczy{\'n}ski}, J., {Ertmer}, W., {Johann}, U., {Penselin}, S., \&
  {Stinner}, P. 1979, Zeitschrift fur Physik A Hadrons and Nuclei, 291, 207

\bibitem[{{Dulick} {et~al.}(2003){Dulick}, {Bauschlicher}, {Burrows}, {Sharp},
  {Ram}, \& {Bernath}}]{Dulick2003}
{Dulick}, M., {Bauschlicher}, Jr., C.~W., {Burrows}, A., {et~al.} 2003, \apj,
  594, 651

\bibitem[{{Duong} {et~al.}(2018){Duong}, {Freeman}, {Asplund}, {Casagrande},
  {Buder}, {Lind}, {Ness}, {Bland-Hawthorn}, {De Silva}, {D'Orazi}, {Kos},
  {Lewis}, {Lin}, {Martell}, {Schlesinger}, {Sharma}, {Simpson}, {Zucker},
  {Zwitter}, {Anguiano}, {Da Costa}, {Hyde}, {Horner}, {Kafle}, {Nataf},
  {Reid}, {Stello}, {Ting}, \& {Wyse}}]{Duong2018}
{Duong}, L., {Freeman}, K.~C., {Asplund}, M., {et~al.} 2018, \mnras, 476, 5216

\bibitem[{{Ekstr{\"o}m} {et~al.}(2008){Ekstr{\"o}m}, {Meynet}, {Chiappini},
  {Hirschi}, \& {Maeder}}]{Ekstrom2008}
{Ekstr{\"o}m}, S., {Meynet}, G., {Chiappini}, C., {Hirschi}, R., \& {Maeder},
  A. 2008, \aap, 489, 685

\bibitem[{{Feuillet} {et~al.}(2018){Feuillet}, {Bovy}, {Holtzman}, {Weinberg},
  {Garc{\'{\i}}a-Hern{\'a}ndez}, {Hearty}, {Majewski}, {Roman-Lopes},
  {Rybizki}, \& {Zamora}}]{Feuillet2018}
{Feuillet}, D.~K., {Bovy}, J., {Holtzman}, J., {et~al.} 2018, \mnras, 477, 2326

\bibitem[{{Fischer} {et~al.}(1967){Fischer}, {H{\"u}hnermann}, \&
  {Kollath}}]{Fischer1967}
{Fischer}, W., {H{\"u}hnermann}, H., \& {Kollath}, K.-J. 1967, Zeitschrift fur
  Physik, 200, 158

\bibitem[{{Frasca} {et~al.}(2018){Frasca}, {Guillout}, {Klutsch}, {Ferrero},
  {Marilli}, {Biazzo}, {Gandolfi}, \& {Montes}}]{Frasca2018}
{Frasca}, A., {Guillout}, P., {Klutsch}, A., {et~al.} 2018, \aap, 612, A96

\bibitem[{{Fuhrmann}(1998)}]{Fuhrmann1998}
{Fuhrmann}, K. 1998, \aap, 338, 161

\bibitem[{{Fuhrmann}(2004)}]{Fuhrmann2004}
{Fuhrmann}, K. 2004, Astronomische Nachrichten, 325, 3

\bibitem[{{Gaia Collaboration} {et~al.}(2018){Gaia Collaboration}, {Brown},
  {Vallenari}, {Prusti}, {de Bruijne}, {Babusiaux}, {Bailer-Jones}, {Biermann},
  {Evans}, {Eyer}, \& et~al.}]{Gaia2018}
{Gaia Collaboration}, {Brown}, A.~G.~A., {Vallenari}, A., {et~al.} 2018, \aap,
  616, A1

\bibitem[{{Gaia Collaboration} {et~al.}(2016){Gaia Collaboration}, {Prusti},
  {de Bruijne}, {Brown}, {Vallenari}, {Babusiaux}, {Bailer-Jones}, {Bastian},
  {Biermann}, {Evans}, \& et~al.}]{Gaia2016}
{Gaia Collaboration}, {Prusti}, T., {de Bruijne}, J.~H.~J., {et~al.} 2016,
  \aap, 595, A1

\bibitem[{{Gehren} {et~al.}(2004){Gehren}, {Liang}, {Shi}, {Zhang}, \&
  {Zhao}}]{Gehren2004}
{Gehren}, T., {Liang}, Y.~C., {Shi}, J.~R., {Zhang}, H.~W., \& {Zhao}, G. 2004,
  \aap, 413, 1045

\bibitem[{{Gehren} {et~al.}(2006){Gehren}, {Shi}, {Zhang}, {Zhao}, \&
  {Korn}}]{Gehren2006}
{Gehren}, T., {Shi}, J.~R., {Zhang}, H.~W., {Zhao}, G., \& {Korn}, A.~J. 2006,
  \aap, 451, 1065

\bibitem[{{Genovali} {et~al.}(2015){Genovali}, {Lemasle}, {da Silva}, {Bono},
  {Fabrizio}, {Bergemann}, {Buonanno}, {Ferraro}, {Fran{\c c}ois}, {Iannicola},
  {Inno}, {Laney}, {Kudritzki}, {Matsunaga}, {Nonino}, {Primas}, {Romaniello},
  {Urbaneja}, \& {Th{\'e}venin}}]{Genovali2015}
{Genovali}, K., {Lemasle}, B., {da Silva}, R., {et~al.} 2015, \aap, 580, A17

\bibitem[{{Gilmore} {et~al.}(2012){Gilmore}, {Randich}, {Asplund}, {Binney},
  {Bonifacio}, {Drew}, {Feltzing}, {Ferguson}, {Jeffries}, {Micela}, \&
  et~al.}]{2012Msngr.147...25G}
{Gilmore}, G., {Randich}, S., {Asplund}, M., {et~al.} 2012, The Messenger, 147,
  25

\bibitem[{{Girardi} {et~al.}(2000){Girardi}, {Bressan}, {Bertelli}, \&
  {Chiosi}}]{Girardi2000}
{Girardi}, L., {Bressan}, A., {Bertelli}, G., \& {Chiosi}, C. 2000, \aaps, 141,
  371

\bibitem[{{Grand} {et~al.}(2018){Grand}, {Bustamante}, {G{\'o}mez}, {Kawata},
  {Marinacci}, {Pakmor}, {Rix}, {Simpson}, {Sparre}, \& {Springel}}]{Grand2018}
{Grand}, R. J.~J., {Bustamante}, S., {G{\'o}mez}, F.~A., {et~al.} 2018, \mnras,
  474, 3629

\bibitem[{{Grisoni} {et~al.}(2017){Grisoni}, {Spitoni}, {Matteucci},
  {Recio-Blanco}, {de Laverny}, {Hayden}, {Mikolaitis}, \&
  {Worley}}]{Grisoni2017}
{Grisoni}, V., {Spitoni}, E., {Matteucci}, F., {et~al.} 2017, \mnras, 472, 3637

\bibitem[{{Gustafsson} {et~al.}(2008){Gustafsson}, {Edvardsson}, {Eriksson},
  {J{\o}rgensen}, {Nordlund}, \& {Plez}}]{Gustafsson2008}
{Gustafsson}, B., {Edvardsson}, B., {Eriksson}, K., {et~al.} 2008, \aap, 486,
  951

\bibitem[{{Handrich} {et~al.}(1969){Handrich}, {Steudel}, \&
  {Walther}}]{Handrich1969}
{Handrich}, E., {Steudel}, A., \& {Walther}, H. 1969, Physics Letters A, 29,
  486

\bibitem[{{Haywood} {et~al.}(2013){Haywood}, {Di Matteo}, {Lehnert}, {Katz}, \&
  {G{\'o}mez}}]{Haywood2013}
{Haywood}, M., {Di Matteo}, P., {Lehnert}, M.~D., {Katz}, D., \& {G{\'o}mez},
  A. 2013, \aap, 560, A109

\bibitem[{{Heiter} {et~al.}(2015){Heiter}, {Lind}, {Asplund}, {Barklem},
  {Bergemann}, {Magrini}, {Masseron}, {Mikolaitis}, {Pickering}, \&
  {Ruffoni}}]{Heiter2015}
{Heiter}, U., {Lind}, K., {Asplund}, M., {et~al.} 2015, \physscr, 90, 054010

\bibitem[{{Hermann} {et~al.}(1993){Hermann}, {Lasnitschka}, {Schwabe}, \&
  {Spengler}}]{Hermann1993}
{Hermann}, G., {Lasnitschka}, G., {Schwabe}, C., \& {Spengler}, D. 1993,
  Spectrochimica Acta, 48, 1259

\bibitem[{{Johann} {et~al.}(1981){Johann}, {Dembczy{\'n}ski}, \&
  {Ertmer}}]{Johann1981}
{Johann}, U., {Dembczy{\'n}ski}, J., \& {Ertmer}, W. 1981, Zeitschrift fur
  Physik A Hadrons and Nuclei, 303, 7

\bibitem[{{Joshi}(2007)}]{Joshi07}
{Joshi}, Y.~C. 2007, \mnras, 378, 768

\bibitem[{{Jurgenson} {et~al.}(2016){Jurgenson}, {Fischer}, {McCracken},
  {Sawyer}, {Giguere}, {Szymkowiak}, {Santoro}, \& {Muller}}]{Jurgenson2016}
{Jurgenson}, C., {Fischer}, D., {McCracken}, T., {et~al.} 2016, Journal of
  Astronomical Instrumentation, 5, 1650003

\bibitem[{{Karakas}(2010)}]{Karakas2010}
{Karakas}, A.~I. 2010, \mnras, 403, 1413

\bibitem[{{Katz} {et~al.}(2019){Katz}, {Sartoretti}, {Cropper}, {Panuzzo},
  {Seabroke}, {Viala}, {Benson}, {Blomme}, {Jasniewicz}, {Jean-Antoine},
  {Huckle}, {Smith}, {Baker}, {Crifo}, {Damerdji}, {David}, {Dolding},
  {Fr{\'e}mat}, {Gosset}, {Guerrier}, {Guy}, {Haigron}, {Jan{\ss}en},
  {Marchal}, {Plum}, {Soubiran}, {Th{\'e}venin}, {Ajaj}, {Allende Prieto},
  {Babusiaux}, {Boudreault}, {Chemin}, {Delle Luche}, {Fabre}, {Gueguen},
  {Hambly}, {Lasne}, {Meynadier}, {Pailler}, {Panem}, {Royer}, {Tauran},
  {Zurbach}, {Zwitter}, {Arenou}, {Bossini}, {Gerssen}, {G{\'o}mez},
  {Lemaitre}, {Leclerc}, {Morel}, {Munari}, {Turon}, {Vallenari}, \& {{\v
  Z}erjal}}]{Katz2019}
{Katz}, D., {Sartoretti}, P., {Cropper}, M., {et~al.} 2019, \aap, 622, A205

\bibitem[{{Kobayashi} \& {Nakasato}(2011)}]{Kobayashi2011}
{Kobayashi}, C. \& {Nakasato}, N. 2011, \apj, 729, 16

\bibitem[{{Kobayashi} {et~al.}(2006){Kobayashi}, {Umeda}, {Nomoto}, {Tominaga},
  \& {Ohkubo}}]{Kobayashi2006}
{Kobayashi}, C., {Umeda}, H., {Nomoto}, K., {Tominaga}, N., \& {Ohkubo}, T.
  2006, \apj, 653, 1145

\bibitem[{{Kordopatis} {et~al.}(2011){Kordopatis}, {Recio-Blanco}, {de
  Laverny}, {Gilmore}, {Hill}, {Wyse}, {Helmi}, {Bijaoui}, {Zoccali}, \&
  {Bienaym{\'e}}}]{Kordopatis2011}
{Kordopatis}, G., {Recio-Blanco}, A., {de Laverny}, P., {et~al.} 2011, \aap,
  535, A107

\bibitem[{{Korotin} {et~al.}(2017){Korotin}, {Andrievsky}, {Caffau}, \&
  {Bonifacio}}]{Korotin2017}
{Korotin}, S., {Andrievsky}, S., {Caffau}, E., \& {Bonifacio}, P. 2017, in
  Astronomical Society of the Pacific Conference Series, Vol. 510, Stars: From
  Collapse to Collapse, ed. Y.~Y. {Balega}, D.~O. {Kudryavtsev}, I.~I.
  {Romanyuk}, \& I.~A. {Yakunin}, 141

\bibitem[{{Kurucz}(1993)}]{Kurucz1993}
{Kurucz}, R. 1993, Diatomic Molecular Data for Opacity Calculations.~Kurucz
  CD-ROM No.~15.~Cambridge, Mass.: Smithsonian Astrophysical Observatory,
  1993., 15

\bibitem[{{Li} {et~al.}(2018){Li}, {Zhao}, {Zhai}, \& {Jia}}]{Li2018}
{Li}, C., {Zhao}, G., {Zhai}, M., \& {Jia}, Y. 2018, \apj, 860, 53

\bibitem[{{Luri} {et~al.}(2018){Luri}, {Brown}, {Sarro}, {Arenou},
  {Bailer-Jones}, {Castro-Ginard}, {de Bruijne}, {Prusti}, {Babusiaux}, \&
  {Delgado}}]{Luri2018}
{Luri}, X., {Brown}, A.~G.~A., {Sarro}, L.~M., {et~al.} 2018, \aap, 616, A9

\bibitem[{{Majewski} {et~al.}(2017){Majewski}, {Schiavon}, {Frinchaboy},
  {Allende Prieto}, {Barkhouser}, {Bizyaev}, {Blank}, {Brunner}, {Burton},
  {Carrera}, {Chojnowski}, {Cunha}, {Epstein}, {Fitzgerald}, {Garc{\'{\i}}a
  P{\'e}rez}, {Hearty}, {Henderson}, {Holtzman}, {Johnson}, {Lam}, {Lawler},
  {Maseman}, {M{\'e}sz{\'a}ros}, {Nelson}, {Nguyen}, {Nidever}, {Pinsonneault},
  {Shetrone}, {Smee}, {Smith}, {Stolberg}, {Skrutskie}, {Walker}, {Wilson},
  {Zasowski}, {Anders}, {Basu}, {Beland}, {Blanton}, {Bovy}, {Brownstein},
  {Carlberg}, {Chaplin}, {Chiappini}, {Eisenstein}, {Elsworth}, {Feuillet},
  {Fleming}, {Galbraith-Frew}, {Garc{\'{\i}}a}, {Garc{\'{\i}}a-Hern{\'a}ndez},
  {Gillespie}, {Girardi}, {Gunn}, {Hasselquist}, {Hayden}, {Hekker}, {Ivans},
  {Kinemuchi}, {Klaene}, {Mahadevan}, {Mathur}, {Mosser}, {Muna}, {Munn},
  {Nichol}, {O'Connell}, {Parejko}, {Robin}, {Rocha-Pinto}, {Schultheis},
  {Serenelli}, {Shane}, {Silva Aguirre}, {Sobeck}, {Thompson}, {Troup},
  {Weinberg}, \& {Zamora}}]{2017AJ....154...94M}
{Majewski}, S.~R., {Schiavon}, R.~P., {Frinchaboy}, P.~M., {et~al.} 2017, \aj,
  154, 94

\bibitem[{{Masseron} {et~al.}(2014){Masseron}, {Plez}, {Van Eck}, {Colin},
  {Daoutidis}, {Godefroid}, {Coheur}, {Bernath}, {Jorissen}, \&
  {Christlieb}}]{Masseron2014}
{Masseron}, T., {Plez}, B., {Van Eck}, S., {et~al.} 2014, \aap, 571, A47

\bibitem[{{Michalik} {et~al.}(2015){Michalik}, {Lindegren}, \&
  {Hobbs}}]{Michalik15}
{Michalik}, D., {Lindegren}, L., \& {Hobbs}, D. 2015, \aap, 574, A115

\bibitem[{{Miglio} {et~al.}(2017){Miglio}, {Chiappini}, {Mosser}, {Davies},
  {Freeman}, {Girardi}, {Jofr{\'e}}, {Kawata}, {Rendle}, {Valentini},
  {Casagrande}, {Chaplin}, {Gilmore}, {Hawkins}, {Holl}, {Appourchaux},
  {Belkacem}, {Bossini}, {Brogaard}, {Goupil}, {Montalb{\'a}n}, {Noels},
  {Anders}, {Rodrigues}, {Piotto}, {Pollacco}, {Rauer}, {Allende Prieto},
  {Avelino}, {Babusiaux}, {Barban}, {Barbuy}, {Basu}, {Baudin}, {Benomar},
  {Bienaym{\'e}}, {Binney}, {Bland-Hawthorn}, {Bressan}, {Cacciari},
  {Campante}, {Cassisi}, {Christensen-Dalsgaard}, {Combes}, {Creevey}, {Cunha},
  {Jong}, {Laverny}, {Degl'Innocenti}, {Deheuvels}, {Depagne}, {Ridder}, {Di
  Matteo}, {Di Mauro}, {Dupret}, {Eggenberger}, {Elsworth}, {Famaey},
  {Feltzing}, {Garc{\'{\i}}a}, {Gerhard}, {Gibson}, {Gizon}, {Haywood},
  {Handberg}, {Heiter}, {Hekker}, {Huber}, {Ibata}, {Katz}, {Kawaler},
  {Kjeldsen}, {Kurtz}, {Lagarde}, {Lebreton}, {Lund}, {Majewski}, {Marigo},
  {Martig}, {Mathur}, {Minchev}, {Morel}, {Ortolani}, {Pinsonneault}, {Plez},
  {Prada Moroni}, {Pricopi}, {Recio-Blanco}, {Reyl{\'e}}, {Robin}, {Roxburgh},
  {Salaris}, {Santiago}, {Schiavon}, {Serenelli}, {Sharma}, {Silva Aguirre},
  {Soubiran}, {Steinmetz}, {Stello}, {Strassmeier}, {Ventura}, {Ventura},
  {Walton}, \& {Worley}}]{Miglio2017}
{Miglio}, A., {Chiappini}, C., {Mosser}, B., {et~al.} 2017, Astronomische
  Nachrichten, 338, 644

\bibitem[{{Mikolaitis} {et~al.}(2017){Mikolaitis}, {de Laverny},
  {Recio-Blanco}, {Hill}, {Worley}, \& {de Pascale}}]{Mikolaitis2017}
{Mikolaitis}, {\v S}., {de Laverny}, P., {Recio-Blanco}, A., {et~al.} 2017,
  \aap, 600, A22

\bibitem[{{Mikolaitis} {et~al.}(2014){Mikolaitis}, {Hill}, {Recio-Blanco}, {de
  Laverny}, {Allende Prieto}, {Kordopatis}, {Tautvai{\v s}iene}, {Romano},
  {Gilmore}, {Randich}, {Feltzing}, {Micela}, {Vallenari}, {Alfaro}, {Bensby},
  {Bragaglia}, {Flaccomio}, {Lanzafame}, {Pancino}, {Smiljanic}, {Bergemann},
  {Carraro}, {Costado}, {Damiani}, {Hourihane}, {Jofr{\'e}}, {Lardo},
  {Magrini}, {Maiorca}, {Morbidelli}, {Sbordone}, {Sousa}, {Worley}, \&
  {Zaggia}}]{Mikolaitis2014}
{Mikolaitis}, {\v S}., {Hill}, V., {Recio-Blanco}, A., {et~al.} 2014, \aap,
  572, A33

\bibitem[{{Mikolaitis} {et~al.}(2018){Mikolaitis}, {Tautvai{\v s}ien{\.e}},
  {Drazdauskas}, {Minkevi{\v c}i{\= u}t{\.e}}, {Klebonas}, {Bagdonas}, {Pak{\v
  s}ien{\.e}}, \& {Janulis}}]{Mikolaitis2018}
{Mikolaitis}, {\v S}., {Tautvai{\v s}ien{\.e}}, G., {Drazdauskas}, A., {et~al.}
  2018, \pasp, 130, 074202

\bibitem[{{Mints} \& {Hekker}(2017)}]{Mints17}
{Mints}, A. \& {Hekker}, S. 2017, \aap, 604, A108

\bibitem[{{Mishenina} {et~al.}(2015){Mishenina}, {Gorbaneva}, {Pignatari},
  {Thielemann}, \& {Korotin}}]{Mishenina2015}
{Mishenina}, T., {Gorbaneva}, T., {Pignatari}, M., {Thielemann}, F.-K., \&
  {Korotin}, S.~A. 2015, \mnras, 454, 1585

\bibitem[{{Mishenina} {et~al.}(2017){Mishenina}, {Pignatari}, {C{\^o}t{\'e}},
  {Thielemann}, {Soubiran}, {Basak}, {Gorbaneva}, {Korotin}, {Kovtyukh},
  {Wehmeyer}, {Bisterzo}, {Travaglio}, {Gibson}, {Jordan}, {Paul}, {Ritter},
  {Herwig}, \& {NuGrid Collaboration}}]{Mishenina2017}
{Mishenina}, T., {Pignatari}, M., {C{\^o}t{\'e}}, B., {et~al.} 2017, \mnras,
  469, 4378

\bibitem[{{Mishenina} {et~al.}(2011){Mishenina}, {Gorbaneva}, {Basak},
  {Soubiran}, \& {Kovtyukh}}]{Mishenina2011}
{Mishenina}, T.~V., {Gorbaneva}, T.~I., {Basak}, N.~Y., {Soubiran}, C., \&
  {Kovtyukh}, V.~V. 2011, Astronomy Reports, 55, 689

\bibitem[{{Mitchell} \& {Mohler}(1965)}]{Mitchell1965}
{Mitchell}, Jr., W.~E. \& {Mohler}, O.~C. 1965, \apj, 141, 1126

\bibitem[{{Moore} {et~al.}(1966){Moore}, {Minnaert}, \& {Houtgast}}]{Moore1966}
{Moore}, C.~E., {Minnaert}, M.~G.~J., \& {Houtgast}, J. 1966, {The solar
  spectrum 2935 A to 8770 A}

\bibitem[{{Neves} {et~al.}(2009){Neves}, {Santos}, {Sousa}, {Correia}, \&
  {Israelian}}]{Neves2009}
{Neves}, V., {Santos}, N.~C., {Sousa}, S.~G., {Correia}, A.~C.~M., \&
  {Israelian}, G. 2009, \aap, 497, 563

\bibitem[{{Nissen} {et~al.}(2017){Nissen}, {Silva Aguirre},
  {Christensen-Dalsgaard}, {Collet}, {Grundahl}, \& {Slumstrup}}]{Nissen2017}
{Nissen}, P.~E., {Silva Aguirre}, V., {Christensen-Dalsgaard}, J., {et~al.}
  2017, \aap, 608, A112

\bibitem[{{Nomoto} {et~al.}(2013){Nomoto}, {Kobayashi}, \&
  {Tominaga}}]{Nomoto2013}
{Nomoto}, K., {Kobayashi}, C., \& {Tominaga}, N. 2013, \araa, 51, 457

\bibitem[{{Prantzos} {et~al.}(2018){Prantzos}, {Abia}, {Limongi}, {Chieffi}, \&
  {Cristallo}}]{Prantzos2018}
{Prantzos}, N., {Abia}, C., {Limongi}, M., {Chieffi}, A., \& {Cristallo}, S.
  2018, \mnras, 476, 3432

\bibitem[{{Ram} {et~al.}(2014){Ram}, {Brooke}, {Bernath}, {Sneden}, \&
  {Lucatello}}]{Ram2014}
{Ram}, R.~S., {Brooke}, J.~S.~A., {Bernath}, P.~F., {Sneden}, C., \&
  {Lucatello}, S. 2014, \apjs, 211, 5

\bibitem[{{Rauer} {et~al.}(2016){Rauer}, {Aerts}, {Cabrera}, \& {PLATO
  Team}}]{Rauer2016}
{Rauer}, H., {Aerts}, C., {Cabrera}, J., \& {PLATO Team}. 2016, Astronomische
  Nachrichten, 337, 961

\bibitem[{{Rauer} {et~al.}(2014){Rauer}, {Catala}, {Aerts}, {Appourchaux},
  {Benz}, {Brandeker}, {Christensen-Dalsgaard}, {Deleuil}, {Gizon}, {Goupil},
  {G{\"u}del}, {Janot-Pacheco}, {Mas-Hesse}, {Pagano}, {Piotto}, {Pollacco},
  {Santos}, {Smith}, {Su{\'a}rez}, {Szab{\'o}}, {Udry}, {Adibekyan}, {Alibert},
  {Almenara}, {Amaro-Seoane}, {Eiff}, {Asplund}, {Antonello}, {Barnes},
  {Baudin}, {Belkacem}, {Bergemann}, {Bihain}, {Birch}, {Bonfils}, {Boisse},
  {Bonomo}, {Borsa}, {Brand{\~a}o}, {Brocato}, {Brun}, {Burleigh}, {Burston},
  {Cabrera}, {Cassisi}, {Chaplin}, {Charpinet}, {Chiappini}, {Church},
  {Csizmadia}, {Cunha}, {Damasso}, {Davies}, {Deeg}, {D{\'{\i}}az}, {Dreizler},
  {Dreyer}, {Eggenberger}, {Ehrenreich}, {Eigm{\"u}ller}, {Erikson}, {Farmer},
  {Feltzing}, {de Oliveira Fialho}, {Figueira}, {Forveille}, {Fridlund},
  {Garc{\'{\i}}a}, {Giommi}, {Giuffrida}, {Godolt}, {Gomes da Silva},
  {Granzer}, {Grenfell}, {Grotsch-Noels}, {G{\"u}nther}, {Haswell}, {Hatzes},
  {H{\'e}brard}, {Hekker}, {Helled}, {Heng}, {Jenkins}, {Johansen},
  {Khodachenko}, {Kislyakova}, {Kley}, {Kolb}, {Krivova}, {Kupka}, {Lammer},
  {Lanza}, {Lebreton}, {Magrin}, {Marcos-Arenal}, {Marrese}, {Marques},
  {Martins}, {Mathis}, {Mathur}, {Messina}, {Miglio}, {Montalban}, {Montalto},
  {Monteiro}, {Moradi}, {Moravveji}, {Mordasini}, {Morel}, {Mortier},
  {Nascimbeni}, {Nelson}, {Nielsen}, {Noack}, {Norton}, {Ofir}, {Oshagh},
  {Ouazzani}, {P{\'a}pics}, {Parro}, {Petit}, {Plez}, {Poretti}, {Quirrenbach},
  {Ragazzoni}, {Raimondo}, {Rainer}, {Reese}, {Redmer}, {Reffert},
  {Rojas-Ayala}, {Roxburgh}, {Salmon}, {Santerne}, {Schneider}, {Schou},
  {Schuh}, {Schunker}, {Silva-Valio}, {Silvotti}, {Skillen}, {Snellen}, {Sohl},
  {Sousa}, {Sozzetti}, {Stello}, {Strassmeier}, {{\v S}vanda}, {Szab{\'o}},
  {Tkachenko}, {Valencia}, {Van Grootel}, {Vauclair}, {Ventura}, {Wagner},
  {Walton}, {Weingrill}, {Werner}, {Wheatley}, \& {Zwintz}}]{Rauer2014}
{Rauer}, H., {Catala}, C., {Aerts}, C., {et~al.} 2014, Experimental Astronomy,
  38, 249

\bibitem[{{Recio-Blanco} {et~al.}(2014){Recio-Blanco}, {de Laverny},
  {Kordopatis}, {Helmi}, {Hill}, {Gilmore}, {Wyse}, {Adibekyan}, {Randich},
  {Asplund}, {Feltzing}, {Jeffries}, {Micela}, {Vallenari}, {Alfaro}, {Allende
  Prieto}, {Bensby}, {Bragaglia}, {Flaccomio}, {Koposov}, {Korn}, {Lanzafame},
  {Pancino}, {Smiljanic}, {Jackson}, {Lewis}, {Magrini}, {Morbidelli},
  {Prisinzano}, {Sacco}, {Worley}, {Hourihane}, {Bergemann}, {Costado},
  {Heiter}, {Joffre}, {Lardo}, {Lind}, \& {Maiorca}}]{Recio2014}
{Recio-Blanco}, A., {de Laverny}, P., {Kordopatis}, G., {et~al.} 2014, \aap,
  567, A5

\bibitem[{{Ricker} {et~al.}(2015){Ricker}, {Winn}, {Vanderspek}, {Latham},
  {Bakos}, {Bean}, {Berta-Thompson}, {Brown}, {Buchhave}, {Butler}, {Butler},
  {Chaplin}, {Charbonneau}, {Christensen-Dalsgaard}, {Clampin}, {Deming},
  {Doty}, {De Lee}, {Dressing}, {Dunham}, {Endl}, {Fressin}, {Ge}, {Henning},
  {Holman}, {Howard}, {Ida}, {Jenkins}, {Jernigan}, {Johnson}, {Kaltenegger},
  {Kawai}, {Kjeldsen}, {Laughlin}, {Levine}, {Lin}, {Lissauer}, {MacQueen},
  {Marcy}, {McCullough}, {Morton}, {Narita}, {Paegert}, {Palle}, {Pepe},
  {Pepper}, {Quirrenbach}, {Rinehart}, {Sasselov}, {Sato}, {Seager},
  {Sozzetti}, {Stassun}, {Sullivan}, {Szentgyorgyi}, {Torres}, {Udry}, \&
  {Villasenor}}]{Ricker2015}
{Ricker}, G.~R., {Winn}, J.~N., {Vanderspek}, R., {et~al.} 2015, Journal of
  Astronomical Telescopes, Instruments, and Systems, 1, 014003

\bibitem[{{Rojas-Arriagada} {et~al.}(2016){Rojas-Arriagada}, {Recio-Blanco},
  {de Laverny}, {Schultheis}, {Guiglion}, {Mikolaitis}, {Kordopatis}, {Hill},
  {Gilmore}, {Randich}, {Alfaro}, {Bensby}, {Koposov}, {Costado},
  {Franciosini}, {Hourihane}, {Jofr{\'e}}, {Lardo}, {Lewis}, {Lind}, {Magrini},
  {Monaco}, {Morbidelli}, {Sacco}, {Worley}, {Zaggia}, \&
  {Chiappini}}]{Rojas2016}
{Rojas-Arriagada}, A., {Recio-Blanco}, A., {de Laverny}, P., {et~al.} 2016,
  \aap, 586, A39

\bibitem[{{Romano} {et~al.}(2010){Romano}, {Karakas}, {Tosi}, \&
  {Matteucci}}]{Romano2010}
{Romano}, D., {Karakas}, A.~I., {Tosi}, M., \& {Matteucci}, F. 2010, \aap, 522,
  A32

\bibitem[{{Saito} {et~al.}(2009){Saito}, {Takada-Hidai}, {Honda}, \&
  {Takeda}}]{Saito2009}
{Saito}, Y.-J., {Takada-Hidai}, M., {Honda}, S., \& {Takeda}, Y. 2009, \pasj,
  61, 549

\bibitem[{{Sch{\"o}nrich} {et~al.}(2010){Sch{\"o}nrich}, {Binney}, \&
  {Dehnen}}]{Schonrich10}
{Sch{\"o}nrich}, R., {Binney}, J., \& {Dehnen}, W. 2010, \mnras, 403, 1829

\bibitem[{{Scott} {et~al.}(2015){Scott}, {Asplund}, {Grevesse}, {Bergemann}, \&
  {Sauval}}]{Scott2015}
{Scott}, P., {Asplund}, M., {Grevesse}, N., {Bergemann}, M., \& {Sauval}, A.~J.
  2015, \aap, 573, A26

\bibitem[{{Sharma} {et~al.}(2018){Sharma}, {Stello}, {Buder}, {Kos},
  {Bland-Hawthorn}, {Asplund}, {Duong}, {Lin}, {Lind}, {Ness}, {Huber},
  {Zwitter}, {Traven}, {Hon}, {Kafle}, {Khanna}, {Saddon}, {Anguiano}, {Casey},
  {Freeman}, {Martell}, {De Silva}, {Simpson}, {Wittenmyer}, \&
  {Zucker}}]{Sharma2018}
{Sharma}, S., {Stello}, D., {Buder}, S., {et~al.} 2018, \mnras, 473, 2004

\bibitem[{{Shi} {et~al.}(2014){Shi}, {Gehren}, {Zeng}, {Mashonkina}, \&
  {Zhao}}]{Shi2014}
{Shi}, J.~R., {Gehren}, T., {Zeng}, J.~L., {Mashonkina}, L., \& {Zhao}, G.
  2014, \apj, 782, 80

\bibitem[{{Skrutskie} {et~al.}(2006){Skrutskie}, {Cutri}, {Stiening},
  {Weinberg}, {Schneider}, {Carpenter}, {Beichman}, {Capps}, {Chester},
  {Elias}, {Huchra}, {Liebert}, {Lonsdale}, {Monet}, {Price}, {Seitzer},
  {Jarrett}, {Kirkpatrick}, {Gizis}, {Howard}, {Evans}, {Fowler}, {Fullmer},
  {Hurt}, {Light}, {Kopan}, {Marsh}, {McCallon}, {Tam}, {Van Dyk}, \&
  {Wheelock}}]{Skrutskie06}
{Skrutskie}, M.~F., {Cutri}, R.~M., {Stiening}, R., {et~al.} 2006, \aj, 131,
  1163

\bibitem[{{Smiljanic} {et~al.}(2014){Smiljanic}, {Korn}, {Bergemann}, {Frasca},
  {Magrini}, {Masseron}, {Pancino}, {Ruchti}, {San Roman}, {Sbordone}, {Sousa},
  {Tabernero}, {Tautvai{\v s}ien{\.e}}, {Valentini}, {Weber}, {Worley},
  {Adibekyan}, {Allende Prieto}, {Barisevi{\v c}ius}, {Biazzo},
  {Blanco-Cuaresma}, {Bonifacio}, {Bragaglia}, {Caffau}, {Cantat-Gaudin},
  {Chorniy}, {de Laverny}, {Delgado-Mena}, {Donati}, {Duffau}, {Franciosini},
  {Friel}, {Geisler}, {Gonz{\'a}lez Hern{\'a}ndez}, {Gruyters}, {Guiglion},
  {Hansen}, {Heiter}, {Hill}, {Jacobson}, {Jofre}, {J{\"o}nsson}, {Lanzafame},
  {Lardo}, {Ludwig}, {Maiorca}, {Mikolaitis}, {Montes}, {Morel}, {Mucciarelli},
  {Mu{\~n}oz}, {Nordlander}, {Pasquini}, {Puzeras}, {Recio-Blanco}, {Ryde},
  {Sacco}, {Santos}, {Serenelli}, {Sordo}, {Soubiran}, {Spina}, {Steffen},
  {Vallenari}, {Van Eck}, {Villanova}, {Gilmore}, {Randich}, {Asplund},
  {Binney}, {Drew}, {Feltzing}, {Ferguson}, {Jeffries}, {Micela}, {Negueruela},
  {Prusti}, {Rix}, {Alfaro}, {Babusiaux}, {Bensby}, {Blomme}, {Flaccomio},
  {Fran{\c c}ois}, {Irwin}, {Koposov}, {Walton}, {Bayo}, {Carraro}, {Costado},
  {Damiani}, {Edvardsson}, {Hourihane}, {Jackson}, {Lewis}, {Lind}, {Marconi},
  {Martayan}, {Monaco}, {Morbidelli}, {Prisinzano}, \&
  {Zaggia}}]{Smiljanic2014}
{Smiljanic}, R., {Korn}, A.~J., {Bergemann}, M., {et~al.} 2014, \aap, 570, A122

\bibitem[{{Sneden} {et~al.}(2016){Sneden}, {Cowan}, {Kobayashi}, {Pignatari},
  {Lawler}, {Den Hartog}, \& {Wood}}]{Sneden2016}
{Sneden}, C., {Cowan}, J.~J., {Kobayashi}, C., {et~al.} 2016, \apj, 817, 53

\bibitem[{{Sneden} {et~al.}(2014){Sneden}, {Lucatello}, {Ram}, {Brooke}, \&
  {Bernath}}]{Sneden2014}
{Sneden}, C., {Lucatello}, S., {Ram}, R.~S., {Brooke}, J.~S.~A., \& {Bernath},
  P. 2014, \apjs, 214, 26

\bibitem[{{Sneden}(1973)}]{Sneden1973}
{Sneden}, C.~A. 1973, PhD thesis, THE UNIVERSITY OF TEXAS AT AUSTIN.

\bibitem[{{Spina} {et~al.}(2016){Spina}, {Mel{\'e}ndez}, {Karakas},
  {Ram{\'{\i}}rez}, {Monroe}, {Asplund}, \& {Yong}}]{Spina2016}
{Spina}, L., {Mel{\'e}ndez}, J., {Karakas}, A.~I., {et~al.} 2016, \aap, 593,
  A125

\bibitem[{{Stetson} \& {Pancino}(2008)}]{Stetson2008}
{Stetson}, P.~B. \& {Pancino}, E. 2008, \pasp, 120, 1332

\bibitem[{{Sullivan} {et~al.}(2015){Sullivan}, {Winn}, {Berta-Thompson},
  {Charbonneau}, {Deming}, {Dressing}, {Latham}, {Levine}, {McCullough},
  {Morton}, {Ricker}, {Vanderspek}, \& {Woods}}]{Sullivan2015}
{Sullivan}, P.~W., {Winn}, J.~N., {Berta-Thompson}, Z.~K., {et~al.} 2015, \apj,
  809, 77

\bibitem[{{Takeda} {et~al.}(2009){Takeda}, {Kaneko}, {Matsumoto}, {Oshino},
  {Ito}, \& {Shibuya}}]{Takeda2009}
{Takeda}, Y., {Kaneko}, H., {Matsumoto}, N., {et~al.} 2009, \pasj, 61, 563

\bibitem[{{Takeda} {et~al.}(2002){Takeda}, {Zhao}, {Chen}, {Qiu}, \&
  {Takada-Hidai}}]{Takeda2002}
{Takeda}, Y., {Zhao}, G., {Chen}, Y.-Q., {Qiu}, H.-M., \& {Takada-Hidai}, M.
  2002, \pasj, 54, 275

\bibitem[{{van den Hoek} \& {Groenewegen}(1997)}]{Hoek1997}
{van den Hoek}, L.~B. \& {Groenewegen}, M.~A.~T. 1997, \aaps, 123

\bibitem[{{Woosley} \& {Weaver}(1995)}]{Woosley1995}
{Woosley}, S.~E. \& {Weaver}, T.~A. 1995, \apjs, 101, 181

\bibitem[{{Yan} {et~al.}(2016){Yan}, {Shi}, {Nissen}, \& {Zhao}}]{Yan2016}
{Yan}, H.~L., {Shi}, J.~R., {Nissen}, P.~E., \& {Zhao}, G. 2016, \aap, 585,
  A102

\bibitem[{{Zhao} {et~al.}(2016){Zhao}, {Mashonkina}, {Yan}, {Alexeeva},
  {Kobayashi}, {Pakhomov}, {Shi}, {Sitnova}, {Tan}, {Zhang}, {Zhang}, {Zhou},
  {Bolte}, {Chen}, {Li}, {Liu}, \& {Zhai}}]{Zhao2016}
{Zhao}, G., {Mashonkina}, L., {Yan}, H.~L., {et~al.} 2016, \apj, 833, 225

\bibitem[{{Zucker} {et~al.}(2012){Zucker}, {de Silva}, {Freeman},
  {Bland-Hawthorn}, \& {Hermes Team}}]{Zucker12}
{Zucker}, D.~B., {de Silva}, G., {Freeman}, K., {Bland-Hawthorn}, J., \&
  {Hermes Team}. 2012, in Astronomical Society of the Pacific Conference
  Series, Vol. 458, Galactic Archaeology: Near-Field Cosmology and the
  Formation of the Milky Way, ed. W.~{Aoki}, M.~{Ishigaki}, T.~{Suda},
  T.~{Tsujimoto}, \& N.~{Arimoto}, 421

\end{thebibliography}

%\newpage
\begin{appendix}

\section{}

\onecolumn
\begin{longtable}{lllll} 
\caption{Contents of the online table.}\\ %  with complete information on the analysed stars.}\\
\hline
Col & Label & Units & Explanations \\
\hline
 1 &ID                  &  ---   & Tycho catalogue identification\\
 2 &TESS\_ID             &  ---   & ID in the TESS catalogue \\
 3 &Teff                & K    & Effective temperature\\
 4 &eTeff               &   K    & Error on effective temperature\\
 5 &Logg                &  dex   & Surface gravity\\
 6 &e\_Logg              &  dex   & Error on surface gravity\\
 7 &[Fe/H]              &  ---   & Metallicity \\
 8 &e\_[Fe/H]            &  ---   & Error on metallicity \\
 9 &Vt                  &   km s$^{-1}$  & Microturbulence velocity\\
10 &e\_Vt                &  km s$^{-1}$   & Error on microturbulence velocity\\
11 &Vrad                & km s$^{-1}$ & Radial velocity\\
12 &e\_Vrad              & km s$^{-1}$ & Error on radial velocity \\
13 &Age                 & Gyr  & Age \\
14 &e\_Age               &  Gyr  & Error on age\\
15 & U                  &  km s$^{-1}$  & U velocity\\
16 &e\_U                 &  km s$^{-1}$   & Error on U velocity\\
17 &V                   & km s$^{-1}$    & V velocity \\
18 &e\_V                 & km s$^{-1}$    & Error on V velocity \\
19 &W                   & km s$^{-1}$    & W velocity \\
20 &e\_W                 & km s$^{-1}$     & Error on W velocity\\
21 & d                  & kpc & Distance calculated 1/plx\\
22 & e\_d               & kpc & Error on distance \\
23 &R$_{mean}$           & kpc    & Mean Galactocentric distance\\
24 &e\_R$_{mean}$        & kpc &Error on mean Galactrocentric distance\\
25 &z$_{max}$                & kpc    & Distance from Galactic plane\\
26 &e\_z$_{max}$              & kpc    & Error on distance from Galactic plane\\
27 &{\it{e}}                   & ---    & Orbital eccentricity\\
28 &e\_{\it{e}}                 & ---  & Error on orbital eccentricity\\
29 &TD/D                & --- & Thick-to-thin disc probability ratio\\
30 &[\ion{Na}{i}/\ion{Fe}{i}]    & ---    & Sodium-to-iron ratio \\
31 &e\_[\ion{Na}{i}/\ion{Fe}{i}]  & ---     & Error on sodium-to-iron ratio\\
32 &[\ion{Mg}{i}/\ion{Fe}{i}]    & ---    & Magnesium-to-iron ratio \\
33 &e\_[\ion{Mg}{i}/\ion{Fe}{i}]  & ---     & Error on magnesium-to-iron ratio\\
34 &[\ion{Al}{i}/\ion{Fe}{i}]    & ---    & Aluminium-to-iron ratio \\
35 &e\_[\ion{Al}{i}/\ion{Fe}{i}]  & ---     & Error on aluminium-to-iron ratio\\
36 &[\ion{Si}{i}/\ion{Fe}{i}]    & ---    & Silicon-to-iron ratio \\
37 &e\_[\ion{Si}{i}/\ion{Fe}{i}]  & ---     & Error on silicon-to-iron ratio\\
38 &[\ion{Si}{ii}/\ion{Fe}{i}]    & ---    & Ionised silicon-to-iron ratio \\
39 &e\_[\ion{Si}{ii}/\ion{Fe}{i}]  & ---     & Error on ionised silicon-to-iron ratio\\
40 &[\ion{S}{i}/\ion{Fe}{i}]    & ---    & Sulphur-to-iron ratio \\
41 &e\_[\ion{S}{i}/\ion{Fe}{i}]  & ---     & Error on sulphur-to-iron ratio\\
42 &[\ion{K}{i}/\ion{Fe}{i}]    & ---    & Potassium-to-iron ratio \\
43 &e\_[\ion{K}{i}/\ion{Fe}{i}]  & ---     & Error on potassium-to-iron ratio\\
44 &[\ion{Ca}{i}/\ion{Fe}{i}]    & ---    & Calcium-to-iron ratio \\
45 &e\_[\ion{Ca}{i}/\ion{Fe}{i}]  & ---     & Error on calcium-to-iron ratio\\
46 &[\ion{Ca}{ii}/\ion{Fe}{i}]    & ---    & Ionised calcium-to-iron ratio \\
47 &e\_[\ion{Ca}{ii}/\ion{Fe}{i}]  & ---     & Error on ionised calcium-to-iron ratio\\
48 &[\ion{Sc}{i}/\ion{Fe}{i}]    & ---    & Scandium-to-iron ratio \\
49 &e\_[\ion{Sc}{i}/\ion{Fe}{i}]  & ---     & Error on scandium-to-iron ratio\\
50 &[\ion{Sc}{ii}/\ion{Fe}{i}]    & ---    & Ionised scandium-to-iron ratio \\
51 &e\_[\ion{Sc}{ii}/\ion{Fe}{i}]  & ---     & Error on ionised scandium-to-iron ratio\\
52 &[\ion{Ti}{i}/\ion{Fe}{i}]    & ---    & Titanium-to-iron ratio \\
53 &e\_[\ion{Ti}{i}/\ion{Fe}{i}]  & ---     & Error on titanium-to-iron ratio\\
54 &[\ion{Ti}{ii}/\ion{Fe}{i}]    & ---    & Ionised titanium-to-iron ratio \\
55 &e\_[\ion{Ti}{ii}/\ion{Fe}{i}]  & ---     & Error on ionised titanium-to-iron ratio\\
56 &[\ion{V}{i}/\ion{Fe}{i}]    & ---    & Vanadium-to-iron ratio \\
57 &e\_[\ion{V}{i}/\ion{Fe}{i}]  & ---     & Error on vanadium-to-iron ratio\\
58 &[\ion{Cr}{i}/\ion{Fe}{i}]    & ---    & Chromium-to-iron ratio \\
59 &e\_[\ion{Cr}{i}/\ion{Fe}{i}]  & ---     & Error on chromium-to-iron ratio\\
60 &[\ion{Cr}{ii}/\ion{Fe}{i}]    & ---    & Ionised chromium-to-iron ratio \\
61 &e\_[\ion{Cr}{ii}/\ion{Fe}{i}]  & ---     & Error on ionised chromium-to-iron ratio\\
62 &[\ion{Mn}{i}/\ion{Fe}{i}]    & ---    & Manganese-to-iron ratio \\
63 &e\_[\ion{Mn}{i}/\ion{Fe}{i}]  & ---     & Error on manganese-to-iron ratio\\
62 &[\ion{Co}{i}/\ion{Fe}{i}]    & ---    & Cobalt-to-iron ratio \\
63 &e\_[\ion{Co}{i}/\ion{Fe}{i}]  & ---     & Error on cobalt-to-iron ratio\\
64 &[\ion{Ni}{i}/\ion{Fe}{i}]    & ---    & Nickel-to-iron ratio \\
65 &e\_[\ion{Ni}{i}/\ion{Fe}{i}]  & ---     & Error on nickel-to-iron ratio\\
66 &[\ion{Cu}{i}/\ion{Fe}{i}]    & ---    & Copper-to-iron ratio \\
67 &e\_[\ion{Cu}{i}/\ion{Fe}{i}]  & ---     & Error on copper-to-iron ratio\\
68 &[\ion{Zn}{i}/\ion{Fe}{i}]    & ---    & Zinc-to-iron ratio \\
69 &e\_[\ion{Zn}{i}/\ion{Fe}{i}]  & ---     & Error on zinc-to-iron ratio\\
70 &[\ion{Fe}{i}/H]    & ---    &  Iron abundance\\
71 &e\_[\ion{Fe}{i}/H]  & ---     & Error on iron abundance\\
72 &[\ion{Fe}{ii}/H]    & ---    & Ionised iron abuncance \\
73 &e\_[\ion{Fe}{ii}/H]  & ---     & Error on ionised iron abundance\\
\noalign{\smallskip}
\hline
%\end{tabular}\\
%}
\label{tab:CDS}

\end{longtable}
\clearpage
\twocolumn

%\section{}

  \begin{figure*}[]%[htb]
  \centering
   \advance\leftskip 0cm
   \includegraphics[scale=0.45]{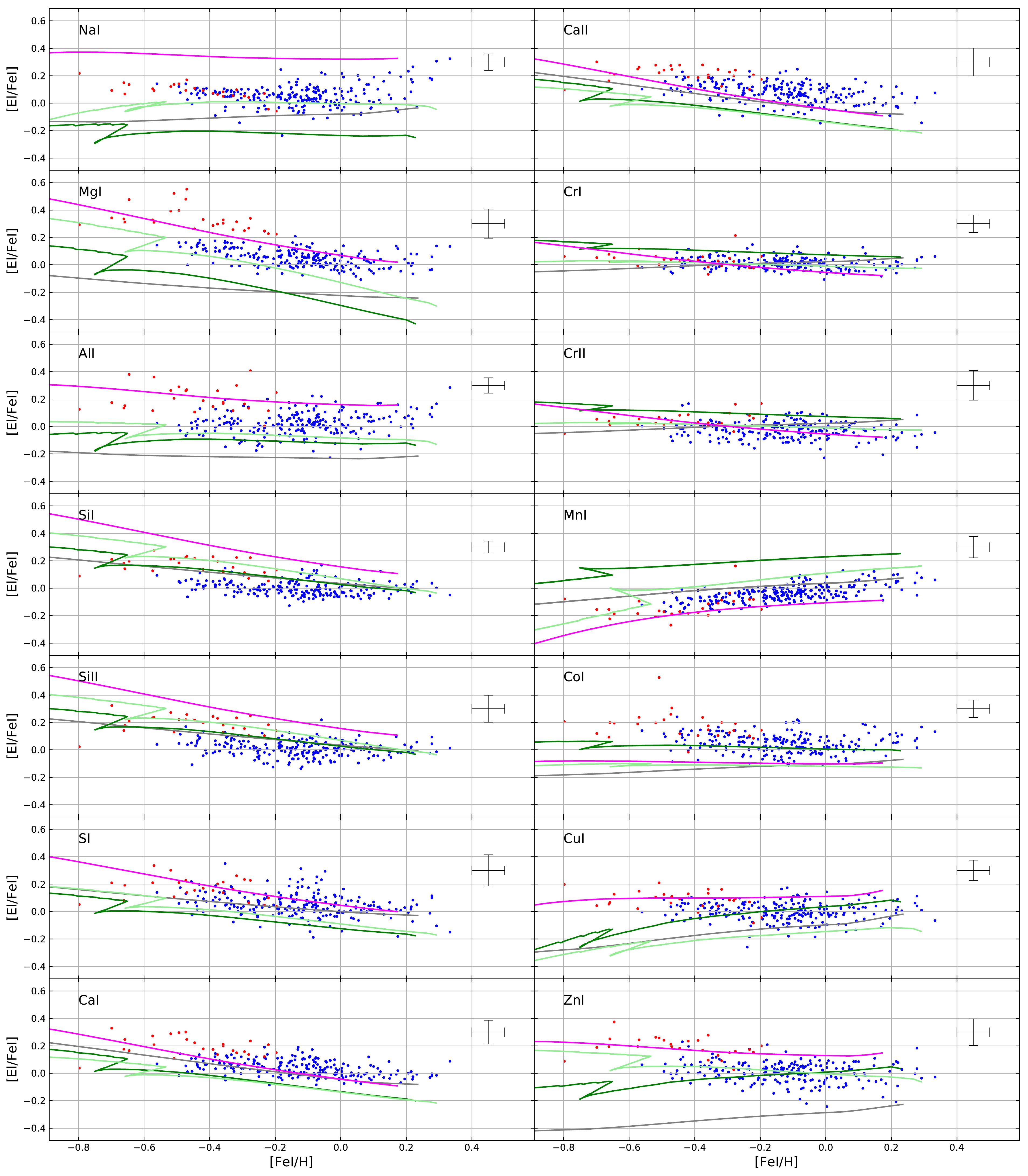}
  \caption{
Observed element-to-iron abundance ratios as a function of the [\ion{Fe}{I}/H] abundance. Shown are cases where some models are able to explain the data. All notations are as in Fig.~\ref{fig:ELEMENTS_separation_failed}
  }
  \label{fig:ELEMENTS_separation}
  \end{figure*}

  \begin{figure*}[htb]
  \centering
   \advance\leftskip 0cm
   \includegraphics[scale=0.45]{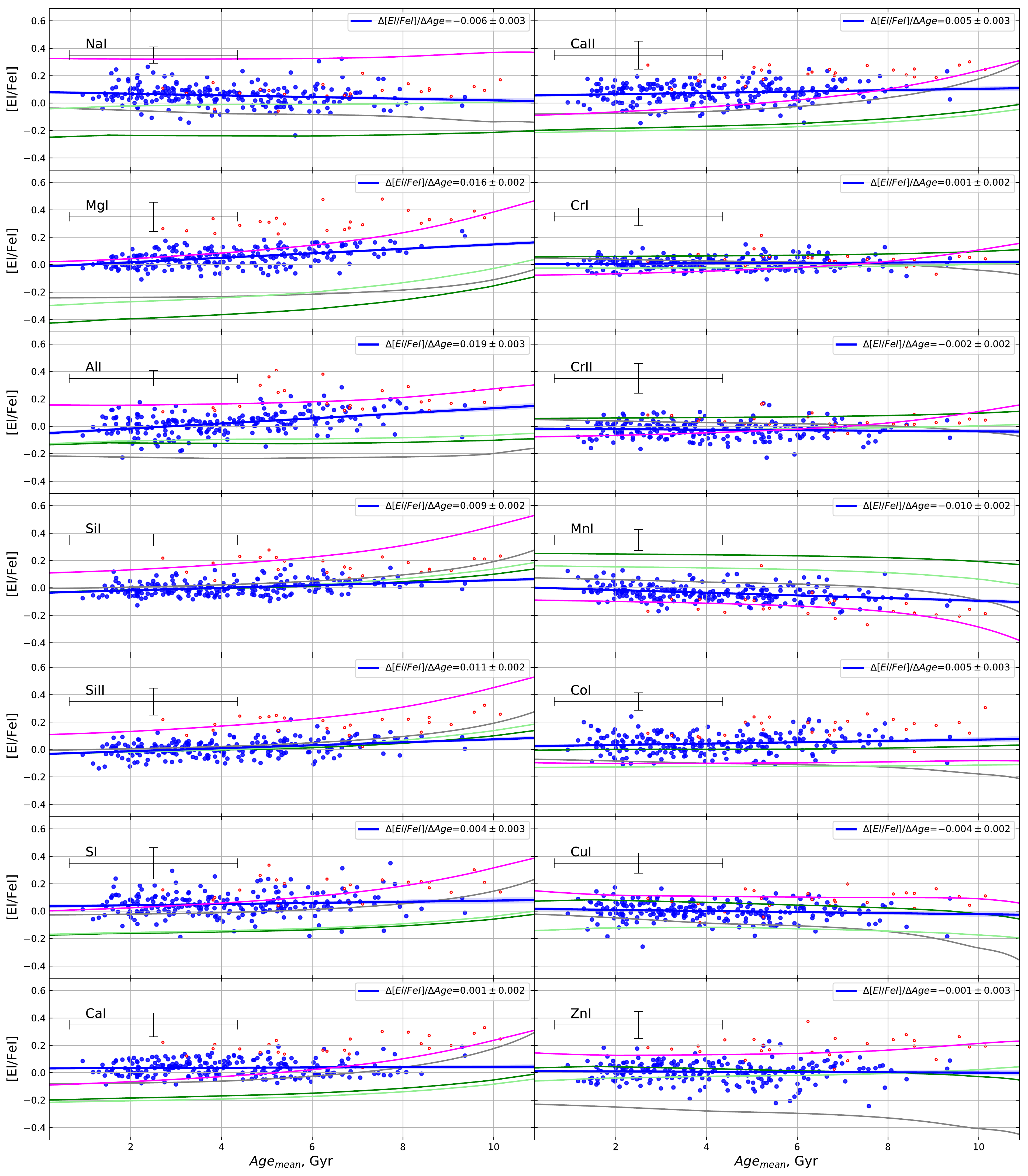}
  \caption{
Observed element-to-iron abundance ratios as a function of the age compared to models. Shown are cases where some models are able to explain the data. All notations are as in Fig.~\ref{fig:ELEMENTS_age_failed}
  }
  \label{fig:ELEMENTS_age}
  \end{figure*}

\end{appendix}

\end{document}